\documentclass{article}

\usepackage[utf8]{inputenc}
\usepackage[english]{babel}
\usepackage[T1]{fontenc}

\usepackage{amsthm}
\usepackage{mathrsfs}
\usepackage{fancyhdr}
\usepackage[colorlinks,linkcolor=red,citecolor=blue,urlcolor=black,filecolor=blue]{hyperref}
\usepackage{graphicx}
\usepackage{amsmath}
\usepackage{amssymb}
\usepackage{dsfont}
\usepackage{mathtools} 

\usepackage{algorithm}
\usepackage{algpseudocode}

\usepackage{natbib}
\usepackage{enumitem}  

\usepackage{authblk}
\usepackage{fullpage}

\usepackage{subcaption} 
\usepackage{booktabs} 

\theoremstyle{plain}
\newtheorem{theorem}{Theorem}[section]

\newtheorem{proposition}[theorem]{Proposition}

\theoremstyle{remark}

\theoremstyle{remark}
\newtheorem{example}{\bf{Example}}[section] 

\newcommand{\prob}{\mathbb{P}}
\newcommand{\mean}{\mathbb{E}}

\newcommand{\Var}{\mathbb{V}}

\newcommand{\cov}{\text{cov}}
\newcommand{\cond}{\,|\,}
\newcommand{\mcond}{\,\middle|\,}
\newcommand{\simiid}{\sim_{\,\text{i.i.d.}}}

\DeclareMathOperator{\Normal}{\mathcal{N}}

\DeclareMathOperator{\Exp}{Exp}

\DeclareMathOperator{\Unif}{\mathcal{U}}

\renewcommand{\epsilon}{\varepsilon}

\newcommand{\ud}{\mathop{}\!\mathrm{d}}

\newcommand{\e}{e} 
\newcommand{\reals}{\mathbb{R}}
\newcommand{\realsp}{\mathbb{R}_{+}}

\newcommand{\Id}{\text{I}}
\newcommand{\T}{^{\top}}

\newcommand{\smallo}{o}
\newcommand{\indic}{\mathds{1}}

\newcommand{\eqdef}{\coloneqq}

\newcommand{\myquad}{\quad\quad}
\newcommand{\iid}{i.i.d.}

\newcommand{\pdf}{\pi}
\newcommand{\pdfapprox}{\widehat{\pdf}}
\newcommand{\pdfabc}{\widehat{\pdf}}
\newcommand{\pdfabcp}{\widehat{\pdf}^{\,\smash{\textnormal{(P)}}}}
\newcommand{\pdfabcf}{\widehat{\pdf}^{\,\smash{\textnormal{(F)}}}}
\newcommand{\pdfabcfa}{\widehat{\pdf}^{\,{\smash{\textnormal{(P')}}}}}
\newcommand{\pdfabcl}{\widehat{\pdf}^{\,\smash{\textnormal{(L)}}}}
\newcommand{\baru}{\bar{u}}

\newcommand{\bary}{\bar{y}}

\newcommand{\barz}{\bar{z}}
\newcommand{\ringy}{\mathring{y}}
\newcommand{\tl}{\tilde}

\newcommand{\Y}{\mathscr{Y}} 
\newcommand{\X}{\mathscr{X}} 
\newcommand{\A}{\mathcal{A}} 
\newcommand{\B}{\mathcal{B}} 

\newcommand{\K}{K} 
\newcommand{\discr}{\Delta} 
\newcommand{\sss}{s} 
\newcommand{\thetatrue}{\theta_{\textnormal{true}}}

\newcommand{\ABCP}{ABC-P} 
\newcommand{\ABCL}{ABC-L} 
\newcommand{\ABCF}{ABC-F} 
\newcommand{\ABCFA}{ABC-P'}

\newcommand{\teta}{\phi} 
\newcommand{\baryteta}{\bary_{\teta}}
\newcommand{\ringyteta}{\ringy_{\teta}}
\newcommand{\mgw}{w} 
\newcommand{\mgu}{u}
\newcommand{\mgy}{y}
\newcommand{\mgv}{v}
\newcommand{\mgx}{x}
\newcommand{\mgwait}{\omega}
\newcommand{\deltat}{\Delta t} 
\newcommand{\yone}{y^{\mathit{1}}}
\newcommand{\ytwo}{y^{\mathit{2}}}
\newcommand{\zonei}{z^{\smash{\mathit{1}(i)}}}
\newcommand{\tp}{\tau} 
\newcommand{\ta}{{\tau_{1}}} 
\newcommand{\tb}{{\tau_{2}}} 
\newcommand{\tc}{{\tau_{3}}} 
\newcommand{\deltameas}{\delta}

\newcommand{\appe}{\textnormal{Supplementary material}}

\allowdisplaybreaks 

\let\originalleft\left
\let\originalright\right
\renewcommand{\left}{\mathopen{}\mathclose\bgroup\originalleft}
\renewcommand{\right}{\aftergroup\egroup\originalright}

\usepackage{color} 
\newcommand{\revcol}{black}

\title{On predictive inference for intractable models via approximate Bayesian computation}

\author[1]{Marko J\"{a}rvenp\"{a}\"{a}\thanks{Corresponding author. Email: \texttt{m.j.jarvenpaa@medisin.uio.no}}}
\author[1,2,3]{Jukka Corander}
\affil[1]{Department of Biostatistics, University of Oslo, Oslo, Norway}
\affil[2]{Department of Mathematics and Statistics, Helsinki Institute of Information Technology (HIIT), University of Helsinki, Helsinki, Finland}
\affil[3]{Wellcome Trust Sanger Institute, Hinxton, Cambridgeshire, United Kingdom}

\date{\today}

\begin{document}

\maketitle

\begin{abstract}
Approximate Bayesian computation (ABC) is commonly used for parameter estimation and model comparison for intractable simulator-based statistical models whose likelihood function cannot be evaluated. In this paper we instead investigate the feasibility of ABC as a generic approximate method for predictive inference, in particular, for computing the posterior predictive distribution of future observations or missing data of interest. We consider three complementary ABC approaches for this goal, each based on different assumptions regarding which predictive density of the intractable model can be sampled from. The case where only simulation from the joint density of the observed and future data given the model parameters can be used for inference is given particular attention and it is shown that the ideal summary statistic in this setting is minimal predictive sufficient instead of merely minimal sufficient (in the ordinary sense). An ABC prediction approach that takes advantage of a certain latent variable representation is also investigated. We additionally show how common ABC sampling algorithms can be used in the predictive settings considered. Our main results are first illustrated by using simple time-series models that facilitate analytical treatment, and later by using two common intractable dynamic models. 

\noindent\textbf{Keywords:} Approximate Bayesian computation, posterior predictive distribution, selection of summary statistics, predictive sufficiency, intractable dynamic models
\end{abstract}

\section{Introduction} \label{sec:intro}

Approximate Bayesian computation has become an important technique for statistical inference in various application fields such as epidemiology and genomics \citep{McKinley2009,Numminen2013,Kypraios2017,Jarvenpaa2019plos}, systems biology \citep{Wilkinson2018,Warne2019} and computational finance \citep{Calvet2014,Frazier2019} to name just a few. 
ABC allows to approximate the posterior distribution of unknown model parameters and the posterior probabilities of models using only forward model simulations. It hence facilitates Bayesian inference in an intuitive -- though usually only in an approximate and computationally intensive -- manner when the likelihood function is intractable in the sense that its analytical form is unavailable, too complex to analyze or too expensive to evaluate. 
Sometimes the terminology of likelihood-free inference (LFI) or simulation-based inference is used for computational methods in this space.  

Despite substantive amount of research on ABC methodology and applications (see e.g.~\citet{Marin2012,Sisson2019,Pesonen2021}) and the fundamental importance of \emph{prediction} in statistical inference, the use of ABC in the context of \emph{predictive inference} has not been widely studied. 
In this paper we investigate ABC as a generic technique to approximate the posterior predictive distribution of some future observations or missing data. 
We denote the unobserved (future or missing) data of interest by $\tl{y}$, the observed data by $y$ and the unknown model parameters by $\theta$. Throughout this paper we assume that the analytical forms of the likelihood function $\pdf(\tl{y},y\cond\theta)$ (and $\pdf(y\cond\theta)$) and the conditional predictive density $\pdf(\tl{y}\cond y,\theta)$ are intractable. 
We especially study a rather general situation where one can jointly simulate from $\pdf(\tl{y},y\cond\theta)$ for each $\theta$ but, importantly, not necessarily from $\pdf(\tl{y}\cond y,\theta)$. 
No specific assumptions of the model structure are made but we also separately analyze such models that facilitate simulation from $\pdf(\tl{y}\cond v,y,\theta)$ where $v$ are unobserved (finite-dimensional) latent variables. 
The goal is to infer the unobserved data $\tl{y}$ 
and possibly also $\theta$ given the observations $y$. 
We consider only an offline setting where data are not obtained sequentially and where summarized data is used as common also in more conventional ABC inference tasks. 

{\color{\revcol}An earlier approach for ABC prediction, used in applied work e.g.~by \citet{McKinley2009,Canale2016,Pesonen2021} and studied theoretically in the context of financial time-series models that facilitate partial analytical treatment by \citet{Frazier2019}, consists of first using ABC to approximate $\pdf(\theta\cond y)$ and then simulating directly from $\pdf(\tl{y}\cond y,\theta)$.} 
However, the latter step can be infeasible or its implementation may require laborious model-specific derivations. 
In contrary, sampling from $\pdf(\tl{y},y\cond\theta)$, in particular, can often be carried out easily -- typically using the same algorithm as sampling from $\pdf(y\cond\theta)$ which is in any case needed for ABC. 
The main advantage of the proposed ABC methods is that they allow predictive inference very generally -- even when the model is specified as a complex computer code that jointly outputs pseudo-data from $\pdf(\tl{y},y\cond\theta)$.
{\color{\revcol}An example would be a temporal Markov model defined via an intractable transition density. If e.g.~only some elements of the last state $y_{\tau}$ were observed, then forward simulation of future data $\tl{y}$ directly via $\pdf(\tl{y}\cond y,\theta) = \pdf(\tl{y}\cond y_{\tau},\theta)$ becomes infeasible. 
Applications include infectious diseases modelling and stochastic biochemical networks, among others, where (semi-)Markov and stochastic differential equation models are used (see e.g.~\citet{Picchini2014,Kypraios2017,Tancredi2019,Wilkinson2018,Buckwar2020}) and where incomplete measurements are often only available. See also \citet{Pesonen2021} where realistic intractable dynamic models are used for prediction, though in a simpler setting as considered here. The focus of this paper is, however, on understanding the predictive ABC approximations and not on particular applications. 
We also offer some new insight on the aforementioned earlier ABC prediction approach and argue that, if applicable, it can be expected to produce more accurate approximation of the posterior predictive distribution than the more generic methods we propose. 
}

{\color{\revcol}Predictive ABC methods are also useful for parameter estimation and model comparison of intractable models. For example, computation of the posterior predictive distribution for the hypothetical future data to be observed given the candidate experimental design and current data is required in sequential Bayesian experimental design. This requires ABC resolution when the model is intractable, see \citet{Hainy2016,Kleinegesse2021}.} 
%
Also, various methods for model assessment and comparison based on data splitting (see e.g.~\citet{Vehtari2012,Burkner2020}), which are yet to be extended to the ABC setting, require the ability to predict data in a hold-out set (or in a cross-validation fold) given the rest of the data. 
%
{\color{\revcol}Predictive inference with tractable models may sometimes require ABC: Bayesian inference could be based on summarized data to make the likelihood function better behaving (see e.g.~\citet{Wood2010,Fasiolo2016}) or to alleviate the effects of possible outlier observations and model misspecification (see e.g.~\citet{Lewis2021}) which can render inference intractable.} 
These potential applications are however not the focus of this work. 

{\color{\revcol}In this paper we only analyze generic ABC prediction methods which are widely applicable, fairly simple to implement and likely easy for a practitioner to understand. It is however worth emphasizing that, obviously, taking advantage of the model structure at hand or its partial analytical tractability, whenever feasible, might lead to more accurate or efficient model-specific algorithms. 
An example would be hidden Markov and state-space models (SSM) with intractable observation models which have been studied by \citet{Jasra2012,Martin2014,Calvet2014,Jasra2015,Vankov2019}. If ABC inference can be based on the full data (instead of summary statistics) as assumed in these works, and if the unobserved latent states are also to be estimated, 
the model-specific ABC algorithms in the above works are presumably most suitable if extended for the prediction task. 
Similar remark holds for Markov jump processes with tractable noise models that in some settings facilitate asymptotically exact particle MCMC methods, see e.g.~\citet{Golightly2011,Wilkinson2018,Warne2020}.
We also consider some SMMs but this is mainly for the simplicity of demonstration. See also \citet{Martin2019} where summary statistics based on auxiliary models are used in ABC parameter estimation with SMMs. 
%
}

The rest of this paper is organized as follows: 
In Section \ref{sec:background} we briefly review the key ideas of Bayesian predictive inference and ABC. 
In Section \ref{sec:main} we first discuss the aforementioned common approach for predictive ABC. We then develop the new ABC prediction approaches and analyze some of their properties. We especially study the selection of summary statistics and observe that these ideally need to be minimal predictive sufficient instead of merely minimal sufficient (in the ordinary sense). In the alternative approach where an assumption of latent variable representation is made, a statistic that is jointly sufficient for both the parameters $\theta$ and the latent variables $v$ relevant for the prediction task is appropriate. 
After that, in Section \ref{sec:abcpredcomp} we show how common ABC samplers such as ABC-MCMC \citep{Marjoram2003} and PMC-ABC \citep{Beaumont2009} can be used for predictive inference and discuss some practical aspects relevant for applications. 
Section \ref{sec:experiments} presents numerical experiments with M/G/1 queue and Lotka-Volterra models which illustrate the main theoretical findings and the quality of the predictive ABC approximations. 
Section \ref{sec:conclusions} summarizes the main results and suggests topics for future work. 
Additional analysis and technical details can be found in the \appe{}.

\section{Background} \label{sec:background}

We denote the observed data as $y=y_{1:n}=(y_1,\ldots,y_n)$ and the unobserved data of interest as $\tl{y}=(\tl{y}_1,\ldots,\tl{y}_{\tl{n}})$. We assume a parametric model $\pdf(\tl{y},y\cond\theta) = \pdf(\tl{y}\cond y,\theta)\pdf(y\cond\theta)$ with a tractable prior density $\pdf(\theta)$ for the unknown model parameters $\theta\in\Theta\subset\reals^{p}$. Unless explicitly stated otherwise, possible latent variables, which we typically denote by $v$ (and $\tl{v}$), are not included to $\theta$ in our notation. 
The individual data points $y_i\in\Y_i\subset\reals^{d_i}$ are typically neither independent nor identically distributed and are associated with known input variables $x_i\in\X_i$ with $x=x_{1:n}=(x_1,\ldots,x_n)$. The same holds for unobserved data $\tl{y}$ with inputs $\tl{x}=(\tl{x}_1,\ldots,\tl{x}_{\tl{n}})$. 
In the case of a time-series model, each $x_i$ would typically denote time and $y_i$ related observation. 
We however suppress the dependence on $x$ and $\tl{x}$ for brevity.

\subsection{Predictive inference} \label{subsec:postpredcomp}

Inference about the unobserved (future) data $\tl{y}$ after observing $y$, which is specifically called predictive inference, is in Bayesian statistics done using the posterior distribution 
\begin{equation}
    \pdf(\tl{y}\cond y) 
    = \int \pdf(\tl{y},\theta\cond y) \ud\theta
    = \int \pdf(\tl{y}\cond\theta,y) \pdf(\theta\cond y) \ud\theta. 
    \label{eq:postpred1}
\end{equation}
This quantity is called the posterior predictive distribution. It captures the uncertainty in $\tl{y}$ both due to the stochastic model and the incomplete knowledge of the model parameters $\theta$. The latter is represented by the posterior distribution 
\begin{equation}
{\color{\revcol}
\pdf(\theta\cond y) 
= \frac{\pdf(y\cond \theta)\pdf(\theta)}{\int \pdf(y\cond \theta) \pdf(\theta) \ud\theta} \propto \pdf(y\cond \theta) \pdf(\theta),}
\label{eq:post}
\end{equation}
where $\pdf(y)=\int \pdf(y\cond \theta) \pdf(\theta) \ud\theta$ is the marginal likelihood. 
As mentioned in \citet[Section~3]{Ohagan2004}, $\theta$ can be seen as a nuisance parameter in predictive inference. However, both $\tl{y}$ and $\theta$ might be of interest in some applications. 

Although we are mainly concerned with prediction in this paper, (\ref{eq:postpred1}) and our proposed ABC methods similarly apply when $\tl{y}$ denotes missing observations to be estimated. 
Furthermore, in some applications latent variables $\tl{v}$ are of the main interest instead of the related future data $\tl{y}$. For example, consider a SSM, where $v=v_{1:n}$ denote the unobserved latent states and $y=y_{1:n}$ the corresponding observations and where the transition and measurement distributions are
\begin{equation}
    \pdf(v_i\cond v_{i-1},\theta), \quad \pdf(y_i\cond v_i,\theta), \quad i=1,\ldots,n,
    \label{eq:ssm}
\end{equation}
respectively. 
The posterior predictive for e.g.~$\tl{v}=v_{n+1}$ is still obtained by applying (\ref{eq:postpred1}) when $\tl{y}$ is replaced by $(\tl{v},\tl{y})$ and $v$ is included to $\theta$ (or marginalized). Hence, we do not handle cases like this separately in what follows. 
When possible, taking the latent variable structure such as (\ref{eq:ssm}) into account can be useful for computational reasons as argued by \citet{Jasra2015} and as considered in Section \ref{subsec:tractcondsimullat} of this paper. 

The posterior predictive distribution (\ref{eq:postpred1}) typically needs to be computed numerically. A common approach is to use {\color{\revcol}the estimator}
\begin{equation}
    \pdf(\tl{y}\cond y) 
    \approx \pdfapprox(\tl{y}\cond y) 
    \eqdef \frac{1}{m}\sum_{j=1}^m \pdf(\tl{y}\cond\theta^{(j)},y),
    \label{eq:postpredapprox1}
\end{equation}
where $\theta^{(j)}, j=1,\ldots,m$, are (possibly correlated) samples from the posterior $\pdf(\theta\cond y)$ obtained e.g.~by some Markov chain Monte Carlo (MCMC) method. If evaluating the conditional predictive density $\pdf(\tl{y}\cond\theta,y)$ needed for (\ref{eq:postpredapprox1}) is difficult, samples from $\pdf(\tl{y}\cond y)$ can be obtained by simulating $\tl{y}^{(j)}$ from $\pdf(\tl{y}\cond\theta^{(j)},y)$ for $j=1,\ldots,m$. 
Theoretical properties of the resulting approximations have recently been extensively studied by \citet{Kruger2020}. 
All in all, provided that one can either evaluate the conditional predictive density or sample from it efficiently, no additional computational challenges arise in principle.

\subsection{Approximate Bayesian computation} \label{subsec:abc}

Before we study predictive inference for intractable models using ABC, we briefly review standard ABC parameter inference. By this we mean the task of approximating the posterior distribution $\pdf(\theta\cond y)$ using only forward model simulations from $\pdf(y\cond\theta)$. More detailed overviews of ABC than given here can be found in \citet{Marin2012,Lintusaari2016,Sisson2019}. 
%
The basic ABC rejection sampler \citep{Tavare1997,Pritchard1999} consists of repeating the following steps for $i=1,\ldots,m$: 
\begin{enumerate}[label=A\arabic*,itemsep=0em]
    \item Simulate $\theta^{(i)}$ from the prior $\pdf(\theta)$, 
    \label{it:abcrej1}
    \item Simulate pseudo-data $z^{(i)}$ from the model given $\theta^{(i)}$,
    \label{it:abcrej2}
    \item Accept $\theta^{(i)}$ if $\discr(\sss(y),\sss(z^{(i)}))\leq h$. 
    \label{it:abcrej3}
\end{enumerate}
The summary statistic\footnote{The function $s$ can depend also on $x$ (and observed data $y$ even when computed for pseudo-data $z$) although this is not explicitly shown in our notation for brevity.}  $\sss:\prod_{i}\Y_i \rightarrow \reals^d$ in \ref*{it:abcrej3} is used to summarize the potentially high-dimensional observed data $y$ and the corresponding simulated data sets $z^{(i)}$. 
The discrepancy function $\discr:\reals^d\times\reals^d\rightarrow\realsp$ and the threshold parameter $h\geq 0$ are further used to compare the similarity of the summary statistics. {\color{\revcol}Commonly $\discr$ is some (weighted) norm $\|\cdot\|$ in $\reals^d$, that is, $\discr(\sss_y,\sss_z)=\|\sss_y-\sss_z\|$, but other choices are also possible.} 
We often denote the observed summary statistic using a short-hand notation $\sss_y\eqdef\sss(y)$ and similarly $\sss_z\eqdef\sss(z)$. 
The accepted samples in \ref*{it:abcrej3} are considered as independent draws from the approximate posterior distribution called the ABC posterior. 

The ABC rejection sampler outlined above and the more advanced samplers discussed in the context of ABC prediction in Section \ref{sec:abcpredcomp} target the ABC posterior 
\begin{equation}
    \pdfabc_{h}(\theta\cond \sss_y) 
    \eqdef \frac{\int \K_h(\discr(\sss_y,\sss_z)) \pdf(\sss_z\cond \theta) \pdf(\theta) \ud \sss_z}{\iint \K_h(\discr(\sss_y,\sss_z)) \pdf(\sss_z\cond \theta) \pdf(\theta) \ud \sss_z \ud \theta}
    \propto \int \K_h(\discr(\sss_y,\sss_z)) \pdf(\sss_z\cond \theta) \pdf(\theta) \ud \sss_z,
    \label{eq:abcapprox1}
\end{equation}
where $\K_h(r) \propto \indic_{r\leq h}$. Other choices of the kernel $\K_h:\realsp\rightarrow\realsp$, such as the Gaussian kernel $\K_h(r) \propto \exp(-r^2/(2h^2))$ where $h> 0$ is a bandwidth parameter, are also possible. 
{\color{\revcol}The intractable likelihood $\pdf(y\cond\theta)$ in (\ref{eq:post}) has essentially been replaced by the ABC likelihood $\pdfabc_{h}(\sss_y\cond\theta) \eqdef \int \K_h(\discr(\sss_y,\sss_z)) \pdf(\sss_z\cond \theta) \ud \sss_z$ to obtain (\ref{eq:abcapprox1}).
%
The quality of this approximation} depends on the selection of the summary statistic $s$ and the threshold (bandwidth) parameter $h$ in particular. In principle $s$ can be the identity mapping which produces no loss of information but, unless the data is low-dimensional, data summarization is necessary for computational efficiency. 
Similarly, the threshold $h$ needs to be selected to balance the tradeoff between approximation accuracy and computational efficiency. 

It is possible to define ABC posterior also via multiple discrepancies $\discr^{(j)}(\sss^{(j)}(y),\sss^{(j)}(z))$, each based on different summary statistic $\sss^{(j)}$ and threshold $h^{(j)}$, by replacing the kernel $\K_h(\discr(\sss_y,\sss_z))$ in (\ref{eq:abcapprox1}) with a product of kernels $\prod_j K_{h^{(j)}}[{\discr^{(j)}(\sss^{(j)}(y),\sss^{(j)}(z))}]$. 
Such an approach with uniform kernels $K_{h^{(j)}}$ has been used by \citet{Pritchard1999,Ratmann2009,Numminen2013} for standard ABC parameter inference. 
The resulting alternative ABC posterior approximation differs from the more common one in that it defines a different acceptance region in the summary statistic space. 
\citet{Wilkinson2013} has shown that these approaches can be interpreted as different measurement error models.

\section{Predictive inference using ABC} \label{sec:main}

In Section \ref{subsec:tractcondsimul} we first consider the situation with tractable $\pdf(\tl{y}\cond y,\theta)$ which typically does not pose significant additional computational or statistical challenges over standard ABC inference. We then analyze a more challenging situation where only joint simulation from $\pdf(\tl{y},y\cond\theta)$ is used (Section \ref{subsec:intrcondsimul}). 
Some special circumstances relevant for all methods are considered in Section \ref{subsec:condsimulnotheta}
and, finally, taking advantage of a latent variable representation is studied in Section \ref{subsec:tractcondsimullat}. 
Table \ref{table:abcpredoverview} shows a summary of the ABC prediction methods. 

\begin{table}[htbp]
\begin{center}
\caption{A simplified overview of the ABC prediction methods considered in this paper. {\color{\revcol}An alternative to \ABCP{} is also briefly analyzed in \appe{} \ref{appe:abcpalt}.} Numerical methods are considered in Section \ref{sec:abcpredcomp}. } \label{table:abcpredoverview}
  \begin{tabular}{lllll}
    \toprule
    Method & Approximation & Sampling feasible from & (Ideal) summary statistic & Details 
    \\
    \midrule
    \ABCF & $\pdfabcf_{h}(\tl{y}\cond y; \sss_y)$, (\ref{eq:abcpostpredf}) & $\pdf(\tl{y}\cond\theta,y)$ and $\pdf(y\cond\theta)$ & Parametric sufficient~(for $\theta$) & Section \ref{subsec:tractcondsimul} 
    \\
    \addlinespace[0.05cm]
    \ABCP & $\pdfabcp_{h}(\tl{y} \cond \sss_y)$, (\ref{eq:abcpostpred}) & $\pdf(\tl{y},y\cond\theta)$ (jointly) & Predictive sufficient (for $\tl{y}$) & Section \ref{subsec:intrcondsimul} 
    \\
    \addlinespace[0.05cm]
    \ABCL & $\pdfabcl_h(\tl{y}\cond y;\sss_y)$, (\ref{eq:lat1}) & $\pdf(\tl{y}\cond y,v,\theta)$ and $\pdf(y,v\cond\theta)$ & Parametric sufficient~(for $\theta$,$v$) & Section \ref{subsec:tractcondsimullat} 
    \\
    \bottomrule
  \end{tabular}
\end{center}
\end{table}

\subsection{Tractable simulation from conditional predictive density} \label{subsec:tractcondsimul}

When the exact posterior $\pdf(\theta\cond y)$ in (\ref{eq:postpred1}) is replaced by the ABC posterior\footnote{Other LFI posterior approximations can also be used and $\pdf(\tl{y}\cond\theta,y)$ can be replaced with  $\pdf(\tl{y}\cond\theta,\sss'_y)$ where $\sss'_y$ is some summary statistic. Sampling from $\pdf(\tl{y}\cond\theta,\sss'_y)$ can however be difficult even if sampling from $\pdf(\tl{y}\cond\theta,y)$ is feasible (and vice versa).}
$\pdfabc_h(\theta\cond \sss_y)$,
the following approximation for the posterior predictive distribution $\pdf(\tl{y}\cond y)$ results:
\begin{equation}
    \pdf(\tl{y}\cond y) 
    \approx
    \pdfabcf_{h}(\tl{y}\cond y; \sss_y) 
    \eqdef \int \pdf(\tl{y}\cond\theta,y) \pdfabc_h(\theta\cond \sss_y) \ud\theta.
    \label{eq:abcpostpredf}
\end{equation}
This intuitive method has been used for applied work by \citet{McKinley2009,Canale2016,Pesonen2021} but without detailed analysis. \citet{Frazier2019} analyzed this approach theoretically and showed that under certain technical conditions the approximate posterior predictive in a sense coincidences with the exact one as the data size becomes arbitrarily large and as $h\rightarrow 0$. Their numerical results interestingly suggest that the approximate posterior predictive can be accurate even if the corresponding ABC posterior of $\theta$ provides a poor approximation. 

In a typical LFI scenario one cannot evaluate $\pdf(\tl{y}\cond y,\theta)$. Samples from $\pdfabcf_{h}(\tl{y}\cond y; \sss_y)$ in (\ref{eq:abcpostpredf}) can be obtained using a similar two stage approach as in Section \ref{subsec:postpredcomp}: One first draws $\theta^{(i)} \sim \pdfabc_h(\theta\cond \sss_y)$ using some standard ABC sampler and finally draws $\tl{y}^{(i)} \sim \pdf(\tl{y}\cond\theta^{(i)},y)$. 
We use the abbreviation \ABCF{} for this approach where ``F''\footnote{This character may also refer to \citet{Frazier2019} who analyzed this approach and used the term \emph{forecasting} while we (interchangeably) use \emph{prediction} in this paper.} informs that this method is directly based on \emph{forward simulation} via $\pdf(\tl{y}\cond y,\theta)$. 
The ABC parameter draws can be recycled for other prediction tasks involving the same model and observed data $y$. 

A limitation of \ABCF{}, mentioned already in Section \ref{sec:intro}, is that sampling from $\pdf(\tl{y}\cond\theta,y)$ may be infeasible when the model is (truly) intractable. \citet{Frazier2019} used \ABCF{} to improve computational efficiency of certain time-series models which facilitate partial analytical treatment and access to $\pdf(\tl{y}\cond\theta,y)$. Also, simulation from $\pdf(\tl{y}\cond\theta,y)$ may be feasible when the model is intractable but is Markovian {\color{\revcol}whose last measured state is fully observed}. 
An important special case where \ABCF{} applies is when $\pdf(\tl{y}\cond y,\theta) = \pdf(\tl{y}\cond\theta)$.  
In particular, the conditional predictive distribution of a model that produces \iid{}~data 
has the same form as the corresponding likelihood function $\pdf(y\cond\theta)$ so that it is in principle possible to even recycle samples generated from $\pdf(y\cond\theta)$ during the ABC sampling from $\pdfabc_h(\theta\cond \sss_y)$. 
A special instance of this is the standard posterior predictive checking of \citet{Gelman1996}, \citet[Section 6.3]{Gelman2013}. In this method \emph{independent} replicated data sets are drawn from the posterior predictive which are then compared against the observed data $y$. This can be implemented using only repeated sampling from $\pdf(y\cond\theta)$ in the ABC case. 

\subsection{Intractable simulation from conditional predictive density} \label{subsec:intrcondsimul}

We now consider the situation where we do not necessary have access to $\pdf(\tl{y}\cond\theta,y)$ but joint simulation from $\pdf(\tl{y},y\cond\theta)$ is feasible. {\color{\revcol}We first write (\ref{eq:postpred1}) as
%
\begin{equation}
    \pdf(\tl{y}\cond y) = \frac{\pdf(\tl{y}, y)}{\pdf(y)}
    = \frac{\int \pdf(\tl{y}, y\cond\theta) \pdf(\theta) \ud\theta}{\iint \pdf(\tl{y}, y\cond\theta) \pdf(\theta) \ud\theta \ud\tl{y}}.
    \label{eq:exactpostpredsimplev2}
\end{equation}
If we then replace $\pdf(\tl{y}, y\cond\theta)$ in (\ref{eq:exactpostpredsimplev2}) with the quantity $\pdfabc_{h}(\tl{y},\sss_y\cond\theta) \eqdef \int \K_h(\discr(\sss_y,\sss_z)) \pdf(\tl{y},\sss_z\cond \theta) \ud \sss_z$, in a similar manner as $\pdf(y\cond\theta)$ is replaced by $\pdfabc_{h}(\sss_y\cond\theta)$ to obtain (\ref{eq:abcapprox1}), the following approximation for the posterior predictive results:}
\begin{align}
\begin{split}
    \pdf(\tl{y}\cond y) 
    \approx \pdfabcp_{h}(\tl{y} \cond \sss_y) 
    &\eqdef \frac{\iint \K_h({\color{\revcol}\discr(\sss_y,\sss_z)}) \pdf(\tl{y}, \sss_z\cond\theta) \pdf(\theta) \ud\sss_z \ud\theta}{\iiint \K_h({\color{\revcol}\discr(\sss_y,\sss_z)}) \pdf(\tl{y}, \sss_z\cond\theta) \pdf(\theta) \ud\sss_z \ud\theta \ud\tl{y}} \\
    &\propto \iint \K_h({\color{\revcol}\discr(\sss_y,\sss_z)}) \pdf(\tl{y}, \sss_z\cond\theta) \pdf(\theta) \ud\sss_z \ud\theta.
    \label{eq:abcpostpred}
\end{split}
\end{align}
We similarly obtain the following joint ABC posterior distribution
\begin{equation}
    \pdf(\tl{y},\theta \cond y) 
    \approx \pdfabcp_{h}(\tl{y},\theta \cond \sss_y) 
    \propto \int \K_h({\color{\revcol}\discr(\sss_y,\sss_z)}) \pdf(\tl{y}, \sss_z\cond\theta) \pdf(\theta) \ud\sss_z. 
    \label{eq:abcjointpostpred}
\end{equation}
{\color{\revcol}These approximations depend on the simulator-based model only via $\pdf(\tl{y}, \sss_z\cond\theta)$ which is the joint distribution of the future (pseudo-)data $\tl{y}$ and the summarized pseudo-data $\sss_z$ given parameter $\theta$. Obviously, we can simulate from this density by first using the simulator-based model to draw from $\pdf(\tl{y},z\cond\theta)$ and then applying the mapping $(\tl{y},z)\mapsto(\tl{y},\sss(z))$.} 

{\color{\revcol}The rest of this section is devoted to the analysis of the approximations (\ref{eq:abcpostpred}) and (\ref{eq:abcjointpostpred}) which we refer as \ABCP{}. A related alternative, based on a ``nested'' ABC approximation of $\pdf(\tl{y}\cond \theta,y)$ in the \ABCF{} approach, is analyzed in \appe{} \ref{appe:abcpalt}. We first observe that marginalizing $\theta$ in (\ref{eq:abcjointpostpred}) naturally produces (\ref{eq:abcpostpred}) while marginalizing $\tl{y}$ reasonably results the standard ABC posterior approximation $\pdfabc_{h}(\theta \cond \sss_y)$. 
We also notice the following connection between \ABCP{} and \ABCF{}:
If we use the basic fact $\pdf(\tl{y}, \sss_z\cond\theta) = \pdf(\tl{y}\cond \sss_z, \theta) \pdf(\sss_z\cond\theta)$ in (\ref{eq:abcpostpred}) and then replace $\pdf(\tl{y}\cond \sss_z, \theta)$ with $\pdf(\tl{y}\cond \sss_y, \theta)$ in the resulting formula} (which is a reasonable approximation in that the integrand of (\ref{eq:abcpostpred}) is approximately zero when $h$ is small and when $\sss_y$ is substantially different from $\sss_z$), we obtain {\color{\revcol}a special case} $\pdfabcf_{h}(\tl{y}\cond \sss_y; \sss_y)$ of the \ABCF{} approximation. {\color{\revcol}Importantly, this connection} additionally implies that the \ABCF{} and \ABCP{} target densities coincide in the case of independent data because then we have $\pdf(\tl{y}\cond \sss_z, \theta) = \pdf(\tl{y}\cond \theta)$.

\subsubsection{Relation to exact posterior predictive distribution} \label{subsec:hlimitanalysis}

We analyze the \ABCP{} target posterior predictive based on the observed summary $\sss_y$ when the uniform kernel $\K_h(r) \propto \indic_{r\leq h}$ is used. We adopt the setting of ABC samplers (Section \ref{subsec:abcpredsampling}) where $h_t$ is a sequence of decreasing thresholds. 
Lebesgue measure is here denoted as $|\cdot|$. 
We make the following assumptions: 
\begin{enumerate}[label=(C\arabic*),itemsep=0em]
    \item The random vector $(\tl{y},\theta,\sss_y)$ has a joint density function $\pdf(\tl{y},\theta,\sss_y)$ with respect to Lebesgue measure, 
    \label{it:a1} 
    \item The acceptance region $\A_t\eqdef\{u\in\reals^d:\discr_t(u,\sss_y)\leq h_t\}$ 
    (or $\A_t\eqdef\{u\in\reals^d:\discr^{(j)}_t(u,\sss_y)\leq h^{(j)}_t \,\forall j\}$ if multiple discrepancies are used) 
    is Lebesgue measurable for each $t$, 
    \label{it:a2}
    \item $\pdf(\sss_y)>0$, 
    \label{it:a3} 
    \item $\lim_{t\rightarrow\infty}|\A_t|=0$, 
    \label{it:a4} 
    \item Sets $\A_t$ have \emph{bounded eccentricity}, that is, for each $\A_t$ there exists a ball $\B_t=\{u\in\reals^d:\|u-\sss_y\|\leq r_t\}$ such that $\A_t\subset\B_t$ and $|\A_t|\geq c|\B_t|$ with some constant $c>0$. \label{it:a5} 
\end{enumerate}
These assumptions are similar to those used by \citet{Prangle2017}, who analyzed the convergence of the ABC posterior of $\theta$, except that we need to consider future data $\tl{y}$ in \ref*{it:a1}. Also, we allow multiple discrepancies in \ref*{it:a2} which is relevant for an approach in Section \ref{subsec:formingdiscr}. The discrepancy $\discr_t$ in \ref*{it:a2} may depend on $t$ although this dependence plays no important role in this paper. 
We refer to \citet[Section~2.3]{Prangle2017} for further discussion on the assumptions which also applies to our predictive ABC setting. 

The following result shows that the \ABCP{} target posterior predictive becomes arbitrarily accurate if the size of the acceptance region shrinks to zero appropriately which is ensured by the above assumptions. The proof of this result is given in \appe{} \ref{appe:theoryproof}. It is easy to verify that \ref*{it:a2}, \ref*{it:a4} and \ref*{it:a5} hold in a typical case where $\discr(\sss_z,\sss_y)$ is some norm and the sequence $h_t$ is such that $h_t\rightarrow 0$ as $t\rightarrow\infty$. 
%
\begin{proposition} \label{thm:abcp}
Assume that a uniform kernel is used and \ref*{it:a1}-\ref*{it:a5} are satisfied. Then, for almost all $(\tl{y},\theta,\sss_y)$ wrt.~the joint density $\pdf(\tl{y},\theta,\sss_y)$, it holds that
\begin{equation}
    \lim_{t\rightarrow\infty} \pdfabcp_{h_t}(\tl{y}\cond\sss_y) = \pdf(\tl{y}\cond\sss_y), \quad 
    \lim_{t\rightarrow\infty} \pdfabcp_{h_t}(\tl{y},\theta\cond\sss_y) = \pdf(\tl{y},\theta\cond\sss_y).
    \label{eq:abcptargetsh0}
\end{equation}
\end{proposition}
We additionally obtain $\pdfabcp_{\infty}(\tl{y}\cond\sss_y) = \pdf(\tl{y}) = \int \pdf(\tl{y}\cond\theta)\pdf(\theta)\ud\theta  = \iint \pdf(\tl{y},y\cond\theta)\pdf(\theta)\ud y\ud\theta $, that is, the ABC posterior predictive coincides with the prior predictive when $h=\infty$. 
{\color{\revcol}A curiosity is that \ABCF{} leads to a different prior predictive density $\pdfabcf_{\infty}(\tl{y}\cond y; \sss_y)=\int\pdf(\tl{y}\cond\theta,y)\pdf(\theta)\ud\theta$ which however clearly agrees with $\pdfabcp_{\infty}(\tl{y}\cond\sss_y)$ if one formally sets $y=\varnothing$ in $\pdfabcf_{\infty}(\tl{y}\cond y; \sss_y)$.}
%
Using some results in \citet{Stein2005,Biau2015} one could likely extend Proposition \ref{thm:abcp} for more general kernel functions. See also \citet{Barber2015} for related analysis. 
In the following we however study the summary statistics selection as this analysis is more relevant for practice and differs from that of the standard ABC inference.

\subsubsection{Selection of summary statistics}

Recall that a statistic $s$ is 
\emph{(classical) sufficient} if the conditional density $\pdf(y\cond \sss_y,\theta)$ does not depend on $\theta$ and 
\emph{Bayes sufficient} if $\pdf(\theta \cond y) = \pdf(\theta \cond \sss_y)$ for each prior density $\pdf(\theta)$ (and for almost all $y$). It can be shown that Bayes sufficiency is equivalent to classical sufficiency when $\theta$ is finite-dimensional, see e.g.~\citet[Theorem~2.14]{Schervish1995}. In what follows we hence consider only Bayes sufficiency\footnote{A technical subtlety is that one could require the condition $\pdf(\theta \cond \sss_y) = \pdf(\theta \cond y)$ to hold only for the prior density actually used for ABC inference instead of all possible choices as in the definition of Bayes sufficiency. In the following analysis we do not need to take a stand which form should hold.} 
which we call \emph{parametric sufficiency} to surely distinguish it from other sufficiency conditions encountered below. 

Under assumptions similar to our \ref*{it:a1}-\ref*{it:a5}, the ABC posterior $\pdfabc_{h_t}(\theta \cond \sss_y)$ converges to $\pdf(\theta \cond \sss_y)$ as $t\rightarrow\infty$, see \citet{Prangle2017} (or \citet[Chapter~1.7]{Sisson2019} for a less precise analysis). 
Now, when the summary statistic $s$ is parametric sufficient, the limiting ABC posterior $\pdf(\theta \cond \sss_y)$ further agrees with the exact posterior $\pdf(\theta \cond y)$. 
The \ABCF{} posterior predictive $\pdfabcf_{h}(\tl{y}\cond y; \sss_y)$ in (\ref{eq:abcpostpredf}) obviously depends on $h$ and $\sss_y$ only via the ABC posterior $\pdfabc_h(\theta\cond \sss_y)$ and hence agrees with the exact posterior predictive $\pdf(\tl{y}\cond y)$ in this scenario. (In Section \ref{subsec:condsimulnotheta} we nonetheless argue that parametric sufficiency is not a necessary condition for this result regarding \ABCF{}.)
However, perhaps surprisingly, parametric sufficiency does not guarantee that an analogous result holds for either \ABCP{} limiting posterior predictive distributions in (\ref{eq:abcptargetsh0}). 

First, we define a summary statistic $s$ to be \emph{predictive sufficient} if
\begin{equation}
    \pdf(\tl{y} \cond y) = \pdf(\tl{y} \cond \sss_y).
    \label{eq:predsuffweak}
\end{equation}
If $\tl{y}$ and $y$ are conditionally independent given $\theta$ (in particular, if the data is \iid{}) so that $\pdf(\tl{y} \cond y, \theta) = \pdf(\tl{y} \cond \theta)$, then parametric sufficiency implies predictive sufficiency (\ref{eq:predsuffweak}) because 
$\pdf(\tl{y} \cond y) = \int \pdf(\tl{y} \cond y,\theta)\pdf(\theta\cond y)\ud\theta = \int \pdf(\tl{y} \cond \theta)\pdf(\theta\cond \sss_y)\ud\theta = \pdf(\tl{y} \cond \sss_y)$. 
This result does not hold more generally as we see later in Example \ref{ex:markovmodel}. 
Classical, Bayes and predictive sufficiency are however all equivalent in a setting of infinitely exchangeable observations of \citet[Section 4.5]{Bernardo1994} but this situation is not particularly relevant for ABC applications. 

When the summary statistic $s$ satisfies the predictive sufficiency condition (\ref{eq:predsuffweak}), it is immediately seen that the \ABCP{} limiting predictive posterior approximation $\pdf(\tl{y}\cond\sss_y)$ of Proposition \ref{thm:abcp} coincides with the exact posterior predictive $\pdf(\tl{y}\cond y)$. The choice $\sss(y)=y$ would trivially satisfy this condition but the summary statistic should be low-dimensional for computational reasons as in standard ABC inference. We conclude that the summary statistic should ideally be minimal predictive sufficient in the sense of (\ref{eq:predsuffweak}) to estimate $\tl{y}$ by using \ABCP{}.

Another, perhaps less known, form of predictive sufficiency exists \citep{Skibinsky1967,Lauritzen1974,Bjornstad1996}. 
This \emph{stronger form of predictive sufficiency} requires the following two conditions:
\begin{align}
    \pdf(\theta\cond y) &= \pdf(\theta\cond \sss_y), \,\text{(that is, } s \text{ is parametric sufficient)}, 
    \label{eq:predsuffstrong1} \\
    \pdf(\tl{y} \cond y, \theta) &= \pdf(\tl{y} \cond \sss_y, \theta).
    \label{eq:predsuffstrong2}
\end{align}
This definition is most natural for our ABC setting but other equivalent definitions exist, see \citet[Section~3.1]{Bjornstad1996}. 
If $s$ is predictive sufficient in the sense of (\ref{eq:predsuffstrong1}) and (\ref{eq:predsuffstrong2}), then $s$ is obviously parametric sufficient but the converse does not hold in general. If $\pdf(\tl{y} \cond \theta, y) = \pdf(\tl{y} \cond \theta)$, then (\ref{eq:predsuffstrong2}) is satisfied and this definition of predictive sufficiency reverts to that of parametric sufficiency. 

When (\ref{eq:predsuffstrong1}) and (\ref{eq:predsuffstrong2}) both hold, we have 
\begin{align}
    \pdf(\tl{y},\theta\cond y) = \pdf(\tl{y}\cond y,\theta)\pdf(\theta\cond y) = \pdf(\tl{y}\cond \sss_y,\theta)\pdf(\theta\cond \sss_y) = \pdf(\tl{y},\theta\cond \sss_y).
    \label{eq:predsuffimpl}
\end{align}
Predictive sufficiency in the sense of (\ref{eq:predsuffstrong1}) and (\ref{eq:predsuffstrong2}) thus implies predictive sufficiency in the sense of (\ref{eq:predsuffweak}). 
This offers a more intuitive way of finding summary statistics for \ABCP{} as compared to (\ref{eq:predsuffweak}): One can take some parametric sufficient statistic and complement it with additional statistics that satisfy (\ref{eq:predsuffstrong2}). This approach may however not produce minimal predictive sufficient statistic in the sense of (\ref{eq:predsuffweak}). 
In practice, where low-dimensional sufficient statistics may not exist and are in any case difficult to find due to the intractability of the model, one can take some ``usual'' summary statistic deemed informative for $\theta$ and try to 
device additional ones that could be informative for prediction in the sense of (\ref{eq:predsuffstrong2}). 

When the summary statistic $\sss$ satisfies (\ref{eq:predsuffimpl}), the joint \ABCP{} limiting approximation $\pdf(\tl{y},\theta\cond\sss_y)$ of Proposition \ref{thm:abcp}, clearly coincides with $\pdf(\tl{y},\theta\cond y)$.
This result is relevant when both $\tl{y}$ and $\theta$ are of interest. 
Of course, $\sss$ should be low-dimensional also in this case.

\begin{example}[Markov model] \label{ex:markovmodel}
This example illustrates the effect of the summary statistics and threshold on the ABC posterior predictive. Consider a discrete-time process 
\begin{equation}
    y_t\cond y_{t-1},\theta \sim \Normal(c+\teta y_{t-1},\sigma^2),
\end{equation}
where $t=1,2,\ldots,n$ with $n>1$ and $y_0=0$. The parameters are $\theta=(c,\teta,\sigma^2)\in\reals\times (-1,1)\times\realsp$ though $\teta$ and $\sigma^2$ are here considered known. 
Consider first the task of estimating $c$ when $y=y_{1:n}=(y_1,\ldots,y_n)\in\reals^n$ is directly observed and the prior for $c$ is $\pdf(c)\propto 1$. 
Then the weighted average $\baryteta$ shown below in (\ref{eq:ex1:cpost}) is a parametric sufficient statistic and when a Gaussian kernel $\Normal(\baryteta\cond \barz_{\teta}, h^2)$ is used, we have 
\begin{equation}
    \pdfabc_h(c\cond \baryteta)
    = \Normal(c\cond\baryteta,\sigma^2/n + h^2), \quad \baryteta \eqdef \frac{(1-\teta)\sum_{i=1}^{n-1}y_i + y_n}{n}. 
    \label{eq:ex1:cpost}
\end{equation}
The exact posterior $\pdf(c\cond y)$ is obtained by setting $h=0$ in (\ref{eq:ex1:cpost}). The derivation of (\ref{eq:ex1:cpost}) and the other results to follow are outlined in \appe{} \ref{appe:markovdetails}. 

We next study \ABCP{} and \ABCF{} posterior predictive densities of $\tl{y}=y_{n+1}$. 
When the Gaussian kernel is again used between the summary statistics under consideration, we obtain\footnote{Deriving an analytical formula for $\pdfabcp_{h}(y_{n+1} \cond y_n, \baryteta)$ is difficult and this quantity presumably has no simple formula when $h>0$ so we consider only the $h=0$ case.
}
\begin{align}
    \pdfabcp_{0}(y_{n+1} \cond y_n,\baryteta) 
    &= 
    \pdf(y_{n+1} \cond y) = \Normal(y_{n+1} \cond \ringyteta, \sigma^2 + \sigma^2/n),
    \quad \ringyteta \eqdef \baryteta + \teta y_n, 
    \label{eq:ex1:postpredtrue} \\
    %
    \pdfabcp_{h}(y_{n+1} \cond \baryteta) 
    &= \Normal\left(y_{n+1} \mcond b_{\teta, n}\baryteta, a_{\teta, n}\sigma^2 + \sigma^2/n + b_{\teta, n}^2h^2\right),
    \label{eq:ex1:postprednonopt} \\
    a_{\teta, n} &\eqdef \frac{1-\teta^{2(n+1)}}{1-\teta^2} - \frac{\teta^2(1-\teta^n)^2}{n(1-\teta)^2} \geq 1, 
    \quad b_{\teta, n} \eqdef 1 + \frac{\teta(1-\teta^n)}{1-\teta} \label{eq:ex1:ab} \\
    \pdfabcp_{h}(y_{n+1} \cond \ringyteta) &= \Normal(y_{n+1} \cond \ringyteta, \sigma^2 + \sigma^2/n + h^2).
    \label{eq:ex1:postpredopt}
\end{align}
While the statistic $\baryteta$ is sufficient for estimating $c$, (\ref{eq:ex1:postprednonopt}) shows that it is neither sufficient for $y_{n+1}$ nor a good choice in practice. Namely, even when $h=0$, the mean prediction of (\ref{eq:ex1:postprednonopt}) can be very different than that of the exact posterior predictive in (\ref{eq:ex1:postpredtrue}). Furthermore, it follows from (\ref{eq:ex1:ab}) that the variance in (\ref{eq:ex1:postprednonopt}) is always larger than that of (\ref{eq:ex1:postpredtrue}) except for some special cases such as the i.i.d.~data case $\phi=0$ where they are equal. 
On the other hand, the predictive \ABCP{} posteriors based on $(\baryteta,y_n)$ and $\ringyteta \eqdef \baryteta + \teta y_n$ in (\ref{eq:ex1:postpredtrue}) and (\ref{eq:ex1:postpredopt}), respectively, both produce the exact posteriors when $h=0$. Both summary statistics are predictive sufficient in the weaker sense. In fact, the former is sufficient for $c$ and $y_{n+p}$ for any $p\geq 1$ while the latter is that for $y_{n+1}$. 
Similarly as with (\ref{eq:ex1:postprednonopt}), one can show that $\ringyteta$ is not a good summary statistic for estimating $c$. 
The emergence of the statistic $y_n$ is not surprising given the Markov property of the model. 

We also compute
\begin{align}
    \begin{split}
    \pdfabcf_{h}(y_{n+1} \cond y; \baryteta) 
    = \pdfabcf_{h}(y_{n+1} \cond y_n; \baryteta) 
    &= \int_{-\infty}^{\infty} \pdf(y_{n+1}\cond y_{n},c) \pdfabc_h(c\cond \baryteta) \ud c \\
    &= \Normal(y_{n+1} \cond \ringyteta, \sigma^2 + \sigma^2/n + h^2),    
    \end{split}
    \label{eq:ex1:postpredf}
\end{align}
which shows that $\pdfabcf_{0}(y_{n+1} \cond y; \baryteta)=\pdf(y_{n+1} \cond y)$, as expected.
We see that even a fairly large bandwidth $h$ may not substantially deteriorate the quality of the \ABCF{} posterior predictive: Heuristically comparing the ABC posterior for $c$ in (\ref{eq:ex1:cpost}) and the \ABCF{} posterior predictive for $y_{n+1}$ in (\ref{eq:ex1:postpredf}) reveals that $h$ needs to be small as compared to $\sigma^2/n$ in the former case whereas as compared to $(1+1/n)\sigma^2 \gg \sigma^2/n$ in the latter case. 
We also see that \ABCF{} in (\ref{eq:ex1:postpredf}) and \ABCP{} based on $\ringyteta$ in (\ref{eq:ex1:postpredopt}) have the same target ABC posterior predictive. These methods are however based on different summary statistics and hence require different choices of $h$ to yield similar computational efficiency in practice. 

Finally, we analyze the summary statistics $s^{(1)}(y) \eqdef \baryteta$, $s^{(2)}(y) \eqdef (\baryteta,y_n)$ and $s^{(3)}(y) \eqdef \ringyteta$ empirically in a more practical setting where the common uniform kernel with threshold $h$ is now used. We consider $n=100$ observations and set $\teta=0.5$ and $\sigma^2=1$. Another choice is considered in \appe{} \ref{appe:experiments}. The true value of $c$ is $1$. 
The inference is carried out using ABC-MCMC with $10^6$ simulations (see Section \ref{sec:abcpredcomp} for details) and thresholds adjusted so that the acceptance probabilities are roughly $10\%$ for all three cases. 

The results in Figure~\ref{fig:markov1} agree with the theoretical analysis above. In particular, we see that summary statistic $s^{(3)}$, which is sufficient only for predicting ${y}_{101}$, does not produce accurate posterior for $c$. 
While both $s^{(1)}$ and $s^{(2)}$ are sufficient for $c$, $s^{(1)}$ produces slightly more accurate approximation which is likely because matching this one-dimensional statistic is computationally more efficient than the two-dimensional $s^{(2)}$. As expected, $s^{(1)}$ produces less accurate posterior predictive at $t=101$ as the other summaries but unexpectedly works fairly well when $t=110$ and $t=200$. The \ABCF{} posterior predictive densities with $s^{(1)}$ and $s^{(2)}$ are not shown for clarity and because they both were essentially the same as the exact posterior. (Statistic $s^{(3)}$ was not considered as it would be a meaningless choice for \ABCF{}.) 

\begin{figure}[htbp]
\centering
\includegraphics[width=0.85\textwidth]{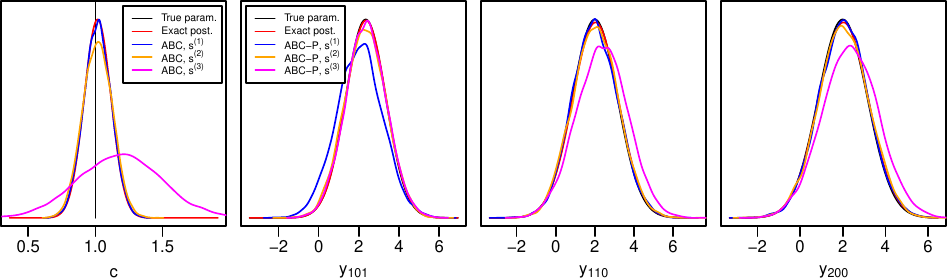}
\caption{Illustration of the effect of summary statistics $s^{(1)}(y)=\baryteta$, $s^{(2)}(y)=(\baryteta,y_n)$ and $s^{(3)}(y)=\ringyteta=\baryteta+\phi y_n$ on the ABC approximation accuracy in Example \ref{ex:markovmodel}. \emph{The first plot} on the left shows the ABC(-P/F) posterior for $c$ and \emph{the three other plots} the \ABCP{} posterior predictive distribution at some future time points. The black vertical line shows the true value of $c$ used to generate the data and the red lines show the exact Gaussian posteriors obtained using (\ref{eq:ex1:cpost}) and (\ref{eq:postpredextrap}) of \appe{} with $h=0$.} \label{fig:markov1}
\end{figure}
\end{example}

{\color{\revcol}
\subsubsection{A remark on computational efficiency} \label{subsubsec:abcpcomp}

Computational efficiency of ABC is often a concern and this is especially the case with \ABCP{}. 
While matching of some ``global'' features of the data may be enough for accurate ABC parameter estimation, \ABCP{} might additionally require matching of some specific ``local'' features. 
Such ``local'' summary statistic for prediction, even if low-dimensional, can have high variability so that a large number of model simulations even with the ``true'' parameter may be needed to obtain a discrepancy evaluation that is small enough to ensure accurate posterior predictive distribution. 
For instance, consider the Markov model in Example \ref{ex:markovmodel} but with the summary statistic $\tl{\sss}(y_{1:n})=y_n$ alone. In \appe{} \ref{appe:predsuffacc}, we show that for any $n\geq 1$, any fixed observations $y_{1:n}$ and threshold $\tl{h}>0$, we have
\begin{equation}
    \max_{c\in\reals}\prob(|z_n-y_n|\leq \tl{h}\cond c) 
    \, \begin{cases} 
    \leq \frac{\sqrt{2}\tl{h}}{\sqrt{\pi\sigma^2}}{\frac{\sqrt{\teta^2-1}}{\sqrt{\teta^{2n}-1}}}, & \textnormal{if } |\teta|>1, \\ 
    \leq \frac{\sqrt{2}\tl{h}}{\sqrt{\pi\sigma^2}\sqrt{n}}, & \textnormal{if } |\teta|=1, \\
    \geq \frac{\sqrt{2}\tl{h}}{\sqrt{\pi\sigma^2}}\sqrt{1-\teta^2}\e^{-\tl{h}^2/(2\sigma^2)}, & \textnormal{if } |\teta|<1,
    \end{cases}
    \label{eq:predsuffacc}
\end{equation}
where $\prob(\cdot)$ is wrt.~simulated data $z$ given parameter $c$. 
It follows from (\ref{eq:predsuffacc}) that the maximum probability of simulating data that matches the observation $y_n$ up to the absolute error $\tl{h}$ goes to zero as $n\rightarrow \infty$ when $|\teta|\geq1$. So, when the data size increases either substantially more simulations are needed or a larger error $\tl{h}$ must be tolerated. The result is unsurprising because the model is a random walk when $|\teta| \geq 1$. 
On the other hand, the maximum acceptance probability is bounded from below with a positive constant in the stationary case $|\teta|<1$. 
}

\subsection{Conditional predictive density with partial parameter dependence} \label{subsec:condsimulnotheta}

We make some remarks on ABC prediction with \ABCP{} and \ABCF{} when the conditional predictive $\pdf(\tl{y}\cond y,\theta)$ depends only partially on $\theta$ and the goal is to estimate only $\tl{y}$. 
We first suppose $\pdf(\tl{y}\cond y,\theta) = \pdf(\tl{y}\cond y)$. 
The first condition of predictive sufficiency (\ref{eq:predsuffstrong1}) then becomes irrelevant and the second condition (\ref{eq:predsuffstrong2}) is the same as the weaker form of predictive sufficiency (\ref{eq:predsuffweak}). Hence, in a (hypothetical) situation where this special condition holds but one can nevertheless only simulate jointly from $\pdf(\tl{y},y\cond\theta)$, \ABCP{} is otherwise implemented as before but the summary statistic used does not need to be informative about $\theta$.  
Additionally, \ABCF{} reverts to direct simulation from $\pdf(\tl{y}\cond y)$ in this case. 
A simple analytical example would be $y_t\cond \theta\simiid \Normal(\theta,1^2)$ for $t=1,\ldots,n$ and $\tl{y}=y_{n+1}$ such that $y_{n+1}\cond y, \theta \sim \Normal(y_{n+1}\cond y_n,1^2)$. Then it clearly holds that $\pdf(\tl{y}\cond y,\theta) = \pdf(\tl{y}\cond y) = \pdf(\tl{y}\cond y_n)$. 
A statistic $\sss$ is called \emph{purely predictive sufficient} in \citet[Section 3.1]{Bjornstad1996} if $\pdf(\tl{y}\cond y,\theta) = \pdf(\tl{y}\cond \sss_y)$ which implies $\pdf(\tl{y}\cond y,\theta) = \pdf(\tl{y}\cond y)$. 
In the example, $y_n$ is purely predictive sufficient. 

We next consider a situation more relevant for practical applications. We write $\theta=(\psi,\eta)$ and suppose $\pdf(\tl{y}\cond y,\theta) = \pdf(\tl{y}\cond y,\psi)$, that is, the conditional predictive density depends only on some of the parameters so that
\begin{equation}
    \pdf(\tl{y}\cond y) = \iint \pdf(\tl{y}\cond y,\psi) \pdf(\psi,\eta\cond y)\ud\psi\ud\eta = \int \pdf(\tl{y}\cond y,\psi) \pdf(\psi\cond y)\ud\psi.
    \label{eq:conddenspartial}
\end{equation}
For instance, $\eta$ could be a parameter of a noise model which is not assumed for future data $\tl{y}$. 
A simple analytical example would be $y_t\cond y_{t-1}, \theta \sim \Normal(\psi+y_{t-1},\eta^2)$ for $t=1,\ldots,n$, $(\psi,\eta)\in\reals\times\realsp$ and $\tl{y}=y_{n+1}$ such that $y_{n+1}\cond y_n,\theta \sim \Normal(y_{n+1}\cond \psi+y_n,1^2)$. 
When (\ref{eq:conddenspartial}) holds, it is in principle enough that the summary statistic $s$ is informative for $\psi$ in \ABCF{}. 
Similarly, in the case of \ABCP{}, the predictive sufficiency conditions (\ref{eq:predsuffstrong1}) and (\ref{eq:predsuffstrong2}) need to concern only $\psi$ instead of $\theta$. In other words, in both cases, $\pdf(\eta\cond\psi, y)$ in principle does not need to be accurately approximated, only $\pdf(\psi\cond y)$ does. 
{\color{\revcol}Design of informative low-dimensional summary statistics for ABC approximation of marginal posteriors, such as $\pdf(\psi\cond y)$ in our above case, has recently been studied by \citet{Drovandi2022}.}

\subsection{Tractable simulation from conditional predictive given latent variables} \label{subsec:tractcondsimullat}

We have now considered ABC prediction when simulation either from both $\pdf(\tl{y}\cond y,\theta)$ and $\pdf(y\cond\theta)$, or jointly from $\pdf(\tl{y},y\cond \theta)$, is feasible. 
In this section we briefly study an alternative that takes advantage of the latent variable representation
\begin{equation}
    \pdf(\tl{y}\cond y, \theta) 
    = \int  \pdf(\tl{y},v\cond y, \theta) \ud v
    = \int \pdf(\tl{y}\cond y, v, \theta) \pdf(v \cond y,\theta) \ud v,
    \label{eq:latentrepresentation}
\end{equation}
where $v$ denotes finite-dimensional, unobserved latent variables involved in the model. 

Combining (\ref{eq:latentrepresentation}) and (\ref{eq:postpred1}) produces
\begin{equation}
    \pdf(\tl{y}\cond y) = \iint \pdf(\tl{y}\cond y, v, \theta) \pdf(v,\theta \cond y) \ud v \ud \theta, \label{eq:latabcpred1}
\end{equation}
which motivates the following approximation of the posterior predictive density: 
\begin{align}
    \pdf(\tl{y}\cond y) \approx \pdfabcl_h(\tl{y}\cond y;\sss_y) 
    &\eqdef \iint \pdf(\tl{y}\cond y, v, \theta) \pdfabc_h(v,\theta\cond \sss_y) \ud v \ud \theta, \label{eq:lat1} \\
    \pdfabc_h(v,\theta\cond \sss_y) &\eqdef \frac{\int \K_h({\color{\revcol}\discr(\sss_y,\sss_z)}) \pdf(\sss_z,v\cond \theta) \pdf(\theta) \ud \sss_z}{\iiint \K_h({\color{\revcol}\discr(\sss_y,\sss_z)}) \pdf(\sss_z,v\cond \theta) \pdf(\theta) \ud \sss_z \ud \theta \ud v}. \label{eq:lat2}
\end{align}
This approach is called \ABCL{} where ``L'' refers to the latent variables. To implement \ABCL{}, the intractable model under consideration must satisfy (\ref{eq:latentrepresentation}) and sampling from both $\pdf(\tl{y}\cond y, v, \theta)$ and $\pdf(y, v\cond\theta)$ must be feasible. 
The latter assumption is evidently weaker than that of \ABCF{}: If one can directly sample from $\pdf(\tl{y}\cond y, \theta)$, then one can also sample from $\pdf(\tl{y}\cond y, v, \theta)$ by formally setting $v=\varnothing$. 
Note that $\pdf(v\cond y,\theta)$ is not assumed tractable -- if one can additionally sample from $\pdf(v\cond y,\theta)$ then one can also sample from $\pdf(\tl{y}\cond y, \theta)$ using (\ref{eq:latentrepresentation}) so that \ABCF{} applies. This is the case in \citet[Section~4]{Frazier2019} where \ABCF{} was used for SMMs. \ABCL{} is in fact similar to \ABCF{} if $\theta$ is allowed to contain latent variables, that is, if $\theta$ is replaced with $(v,\theta)$. 
However, the asymptotic results of \citet{Frazier2019} do not apply for \ABCL{}  as such because the latent variables $v$ cannot typically be consistently estimated. 


One can establish that $\pdfabc_{h_t}(v,\theta\cond \sss_y)\rightarrow \pdf(v,\theta\cond \sss_y)$ as $t\rightarrow\infty$ but we omit the details as they are analogous to Section \ref{subsec:hlimitanalysis} and as a similar result is obtained as a special case of \citet[Proposition~2.1]{Jasra2015}. Clearly, the summary statistic $\sss$ should ideally be jointly sufficient for $v$ and $\theta$ so that 
\begin{equation}
    \pdf(v,\theta\cond y) = \pdf(v,\theta\cond \sss_y). 
    \label{eq:abclsuff}
\end{equation}
This condition resembles Bayes sufficiency. 
Similarly as discussed in Section \ref{subsec:condsimulnotheta}, $\pdf(\tl{y}\cond y, v, \theta)$ may depend only on some components of $(v,\theta)$ so that the sufficiency condition (\ref{eq:abclsuff}) in principle needs to hold only for them. Similarly to the case of \ABCF{}, the \ABCL{} posterior predictive (\ref{eq:lat1}) agrees with the exact posterior predictive $\pdf(\tl{y}\cond y)$ if $\pdfabc_h(v,\theta\cond \sss_y) = \pdf(v,\theta\cond y)$. 

A key difference between \ABCL{} and \ABCP{} is that latter method depends on the full data $y$ only via the summary statistic $\sss_y$ while the former approach, similarly to \ABCF{}, depends directly on $y$ via the conditional predictive $\pdf(\tl{y}\cond y, v, \theta)$ in (\ref{eq:lat1}). 
A potential downside of \ABCL{} is that, even when one could use the fact that the conditional predictive depends only on some components of $(v,\theta)$, ABC inference in high-dimensional space might be needed which is notoriously a hard problem. 

We end this section with a SSM example that demonstrates the selection of summary statistics for \ABCP{} and \ABCL{}. 
Nonetheless, in practice \ABCL{} should be  used for truly intractable models that are different from the basic SSM defined via (\ref{eq:ssm}). Namely, in this case $\pdf(\tl{y}\cond y, v, \theta) = \pdf(\tl{y}\cond v_n, \theta)$ does not depend on $y$ and it can be easily checked that \ABCP{} and \ABCL{} then in fact have the same target ABC posterior predictive. 
%
\begin{example}[State space model] \label{ex:statespacemodel}
Consider the following model 
\begin{equation}
    v_t\cond v_{t-1},\theta \sim \Normal(c+\teta v_{t-1},\sigma^2), \quad y_t\cond v_t,\theta\sim\Normal(v_t,\omega^2),
\end{equation}
where $t=1,2,\ldots,n$ with $n>1$ and $v_0=0$. The set-up is similar to Example \ref{ex:markovmodel} except that the (latent) states are denoted by $v$ and are not observed. Also, $\omega>0$ is an additional fixed parameter. 
We consider only the case $h=0$ and investigate the effect of the summary statistics for estimating $c$ or $\tl{y}=y_{t+1}$ given observations $y=y_{1:n}$. 
Justifications for the various results below are given in \appe{} \ref{appe:ex:statespacemodeldetails}. 

Consider the following scalar-valued summary statistics $s^{(1)}(y_{1:n}) \eqdef \mu\T W^{-1}y_{1:n}$ and $s^{(2)}(y_{1:n}) \eqdef \Sigma_{n:} W^{-1}y_{1:n}$. Here $\mu\in\reals^n$ is such that $\mu_t = (1-\teta^t)/(1-\teta)$, $\Sigma\in\reals^{n\times n}$ is the covariance matrix of $v$ that depends on $\teta$ and $\sigma^2$ (but not on $c$) and whose formula is given in (\ref{eq:markovcov}) of \appe{} \ref{appe:ex:statespacemodeldetails}, $\Sigma_{n:}$ denotes the $n$th row of $\Sigma$ and $W=\Sigma + \omega^2\Id$, where $\Id\in\reals^{n\times n}$ is the identity matrix. 
First of all, $s^{(1)}(y_{1:n})$ is minimal parametric sufficient for $c$ and a linear combination of $s^{(1)}(y_{1:n})$ and $s^{(2)}(y_{1:n})$, shown in (\ref{eq:ssmlinsuffstat}) of \appe{} \ref{appe:ex:statespacemodeldetails}, is minimal predictive sufficient for $y_{n+1}$. 
Hence, these would be ideal summary statistics for computing the ABC posterior of $c$ and the \ABCP{} posterior predictive of $y_{n+1}$, respectively. 

Access to $\pdf(y_{n+1}\cond y_{1:n},v_{1:n},c)$, which here simplifies to $\pdf(y_{n+1}\cond v_n,c)=\Normal(c+\teta v_n,\omega^2+\sigma^2)$, and consequently the ABC posterior of $(v_n,c)$ would be both required for \ABCL{}. Now, $(s^{(1)}(y_{1:n}),s^{(2)}(y_{1:n}))$ qualifies for a sufficient statistic for $(v_n,c)$ in the sense of (\ref{eq:abclsuff}) and is hence sufficient also for $y_{n+1}$ in the \ABCL{} approach. 
This statistic is two-dimensional which is not surprising as it is sufficient for $(v_n,c)\in\reals^2$ in a Gaussian linear system. 
As $\pdf(y_{n+1}\cond y_{1:n},v_{1:n},c)$ does not depend on $y_{1:n}$ and depends on $(v_{1:n},c)$ only via $c+\teta v_n$, it can be argued that the same scalar-valued summary statistic that is predictive sufficient for $y_{n+1}$, is also sufficient for $y_{n+1}$ in the \ABCL{} approach. 
This shows that $(s^{(1)}(y_{1:n}),s^{(2)}(y_{1:n}))$ is not minimal sufficient in this case. 
\end{example}

\section{Using ABC samplers for predictive inference} \label{sec:abcpredcomp}

We next consider numerical methods for \ABCL{} and \ABCP{}.
Our aim is not to develop or advocate some specific algorithm but instead concisely outline how commonly used ABC algorithms can be modified to produce samples from the \ABCL{} and \ABCP{} target densities. We also discuss some practical aspects of predictive ABC inference.

\subsection{Sampling from ABC posterior predictive} \label{subsec:abcpredsampling}

We assume the observed summary statistic $\sss_y$, discrepancy $\discr$, kernel $\K_h$ and threshold $h>0$ are all fixed and consider first the task of sampling from $\pdfabc_h(\theta\cond \sss_y)$ in (\ref{eq:abcapprox1}). 
The ABC target density is augmented to $\pdfabc_h(\sss_z,\theta\cond \sss_y) \propto \K_h({\color{\revcol}\discr(\sss_y,\sss_z)}) \pdf(\sss_z\cond \theta) \pdf(\theta)$ and the proposal density selected as $\pdf(\sss_z\cond\theta)q(\theta)$. Self-normalized importance sampling (IS) can be then used as the resulting (unnormalized) importance weights are free of intractable terms:
\begin{equation}
    \tl{\omega}(\theta) = \frac{\K_h({\color{\revcol}\discr(\sss_y,\sss_z)}) {\pdf(\sss_z\cond \theta)} \pdf(\theta)}{{\pdf(\sss_z\cond \theta)}q(\theta)}
    = \frac{\K_h({\color{\revcol}\discr(\sss_y,\sss_z)}) \pdf(\theta)}{q(\theta)}. 
    \label{eq:abcis}
\end{equation}
Weighted samples $(\tl{\omega}^{(i)},\theta^{(i)})$ from $\pdfabc_h(\theta\cond \sss_y)$ are hence obtained by first sampling $\theta^{(i)}$ from $q(\theta)$, then $z^{(i)}$ from $\pdf(z\cond \theta^{(i)})$, and finally computing $\tl{\omega}^{(i)} = {\K_h({\color{\revcol}\discr(\sss_y,\sss_z^{\smash{(i)}})}) \pdf(\theta^{(i)})} / q(\theta^{(i)})$ with $\sss_z^{\smash{(i)}}=\sss(z^{(i)})$ for $i=1,\ldots,m$. 
Normalized weights ${\omega}^{(i)}$ can be obtained by computing $\omega^{(i)}=\tl{\omega}^{(i)}/\sum_{j=1}^m\tl{\omega}^{(j)}$. 

We similarly define the augmented target density $\pdfabcl_h(\sss_z,v,\theta\cond \sss_y) \propto \K_h({\color{\revcol}\discr(\sss_y,\sss_z)}) \pdf(\sss_z,v\cond \theta) \pdf(\theta)$ for the \ABCL{} approach. If the proposal is analogously set to $\pdf(\sss_z,v\cond\theta)q(\theta)$, the intractable terms $\pdf(\sss_z,v\cond \theta)$ cancel out similarly as $\pdf(\sss_z\cond \theta)$ in (\ref{eq:abcis}) and the same (unnormalized) importance weight $\tl{\omega}(\theta)$ results. 
Hence, weighted samples $(\tl{\omega}^{(i)},\tl{y}^{(i)})$ from $\pdfabcl_h(\tl{y}\cond y;\sss_y)$ in (\ref{eq:lat1}) are obtained by repeating the following steps
\begin{enumerate}[label=L\arabic*,itemsep=0em]
\item Simulate $\theta^{(i)}$ from $q(\theta)$, \label{it:l1}
\item Simulate $(z^{(i)},v^{(i)})$ from $\pdf(z,v\cond \theta^{(i)})$ and compute $\sss_z^{\smash{(i)}}=\sss(z^{(i)})$, \label{it:l2}
\item Compute (unnormalized) IS weight $\tl{\omega}^{(i)} = {\K_h({\color{\revcol}\discr(\sss_y,\sss_z^{\smash{(i)}})}) \pdf(\theta^{(i)})} / q(\theta^{(i)})$, \label{it:l3}
\item Simulate $\tl{y}^{(i)}$ from $\pdf(\tl{y}\cond y, v^{(i)}, \theta^{(i)})$, \label{it:l4}
\end{enumerate}
for $i=1,\ldots,m$. 
Alternatively, a two stage approach can be used where steps \ref*{it:l1}-\ref*{it:l3} are first repeated for all $i$ and \ref*{it:l4} is finally run for those samples with non-zero weights. 
ABC rejection sampler for \ABCL{} results as a special case with $\K_h(r) \propto \indic_{r\leq h}$ and $q(\theta)=\pdf(\theta)$. Up to some caveats discussed later, PMC-ABC \citep{Beaumont2009} and other related sequential IS methods such as \citet{Sisson2007,Toni2009,Moral2012} can be similarly modified to sample from the (augmented) \ABCL{} target. This is also the case with ABC-MCMC \citep{Marjoram2003}. 
All in all, the main modifications are that also $v^{(i)}$ need to be stored (unless e.g.~its weight is zero) and the extra prediction step \ref*{it:l4}. 

Sampling from the \ABCP{} target $\pdfabcp_{h}(\tl{y} \cond \sss_y)$ in (\ref{eq:abcpostpred}) proceeds in a similar fashion as above except that future data $\tl{y}$ is already drawn alongside $z$ in \ref*{it:l2} so that there is no specific prediction step \ref*{it:l4}. 
For a change, we consider ABC-MCMC sampler. 
The augmented \ABCP{} target density is $\pdfabcp_h(\tl{y},\sss_z,\theta\cond \sss_y) \propto \K_h({\color{\revcol}\discr(\sss_y,\sss_z)}) \pdf(\tl{y},\sss_z\cond \theta) \pdf(\theta)$ and the proposal density is $g((\tl{y},\sss_z,\theta),(\tl{y}^*,\sss_z^*,\theta^*)) \eqdef \pdf(\tl{y}^*,\sss_z^*\cond\theta^*)q(\theta^*\cond\theta)$, where $(\tl{y},\sss_z,\theta)$ denotes the current state of the chain and $(\tl{y}^*,\sss_z^*,\theta^*)$ the proposed state. The Metropolis-Hastings acceptance probability for a transition from $(\tl{y},\sss_z,\theta)$ to $(\tl{y}^*,\sss_z^*,\theta^*)$ is then 
\begin{align}
    \begin{split}
    \alpha((\tl{y},\sss_z,\theta),(\tl{y}^*,\sss_z^*,\theta^*)) 
    &= \min\left\{ 1, \frac{\K_h({\color{\revcol}\discr(\sss_y,\sss_z^*)}) \pdf(\tl{y}^*,\sss_z^*\cond \theta^*) \pdf(\theta^*)}{\K_h({\color{\revcol}\discr(\sss_y,\sss_z)}) \pdf(\tl{y},\sss_z\cond \theta) \pdf(\theta)} \frac{g((\tl{y}^*,\sss_z^*,\theta^*),(\tl{y},\sss_z,\theta))}{g((\tl{y},\sss_z,\theta),(\tl{y}^*,\sss_z^*,\theta^*))} \right\} \\
    &= \min\left\{ 1, \frac{\K_h({\color{\revcol}\discr(\sss_y,\sss_z^*)}) \pdf(\theta^*) q(\theta\cond\theta^*)}{\K_h({\color{\revcol}\discr(\sss_y,\sss_z)}) \pdf(\theta) q(\theta^*\cond\theta)} \right\}.
    \end{split}
    \label{eq:mhacc}
\end{align}
Starting from an initial point $(\sss_z^{\smash{(0)}},\theta^{(0)})$ (which should be such that $\K_h({\color{\revcol}\discr(\sss_y,\sss_z^{\smash{(0)}})}) \pdf(\theta^{(0)})>0$), correlated samples from $\pdfabcp_{h}(\tl{y},\theta \cond \sss_y)$ are generated by iterating the following steps for $i=1,\ldots,m$:
\begin{enumerate}[label=P\arabic*,itemsep=0em]
\item Simulate $\theta^*$ from $q(\theta\cond\theta^{(i-1)})$, \label{it:p1}
\item Simulate $(\tl{y}^*,z^*)$ from $\pdf(\tl{y},z\cond \theta^*)$ and compute $\sss_z^*=\sss(z^*)$, \label{it:p2}
\item Set $(\tl{y}^{(i)},z^{(i)},\theta^{(i)}) = (\tl{y}^*,z^*,\theta^*)$ with probability $\alpha((\tl{y}^{(i-1)},\sss_z^{\smash{(i-1)}},\theta^{(i-1)}),(\tl{y}^*,\sss_z^*,\theta^*))$ using (\ref{eq:mhacc}) and otherwise set $(\tl{y}^{(i)},z^{(i)},\theta^{(i)}) = (\tl{y}^{(i-1)},z^{(i-1)},\theta^{(i-1)})$. \label{it:p3}
\end{enumerate}
Analogously to standard ABC-MCMC with the augmented target density $\pdfabc_h(\sss_z,\theta\cond \sss_y)$, the detailed balance condition can be shown to hold so that $\pdfabcp_h(\tl{y},\sss_z,\theta\cond \sss_y)$ is indeed the stationary distribution of the chain generated by the above algorithm. 

The acceptance probability (\ref{eq:mhacc}) and the corresponding IS weight do not depend on $\tl{y}$. Consequently, the computation time of \ABCP{} can be reduced using the latent variable representation
\begin{equation}
    \pdf(\tl{y},z\cond \theta) = \int \pdf(\tl{y}\cond z, v, \theta) \pdf(v,z \cond\theta) \ud v,
    \label{eq:plat}
\end{equation}
which is the same condition as (\ref{eq:latentrepresentation}) assumed for \ABCL{}. 
Specifically, one then first simulates $(z^{(i)},v^{(i)},\theta^{(i)})$ for each $i$ and afterwards simulates only those of the corresponding predictions $\tl{y}^{(i)}$ using $\pdf(\tl{y}\cond z^{(i)}, v^{(i)}, \theta^{(i)})$ that are required to form the final estimate of $\pdf(\tl{y}\cond y)$. For example, those $\tl{y}^{(i)}$ that correspond to ``burn-in'' in the ABC-MCMC sampler or that either have zero IS weight or do not belong to the last population in the PMC-ABC algorithm, are not required. 
In some cases jointly sampling $\tl{y}$ and $z$ is not much slower than sampling only $z$. The above approach can still be useful because the samples can be reused for other related prediction tasks for which the selected summary statistic is deemed informative. 

Adapting some special ABC samplers for \ABCL{} and \ABCP{} may require additional modifications. 
Examples include the adaptive method for adjusting the discrepancy by \citet{Prangle2017}, which is further discussed in Section \ref{subsec:formingdiscr}, and the adaptive threshold selection method by \citet{Simola2021}, which could possibly be modified to account for the posterior predictive at least when $\tl{y}$ is low-dimensional. 

\subsection{On forming the discrepancy for predictive ABC inference} \label{subsec:formingdiscr}

A common choice for the discrepancy is the weighted Euclidean distance between the summary statistics. Its weights are often based on the variability of the summary statistic which is estimated using pilot simulations (or adjusted adaptively as in~\citet{Prangle2017}). 
However, such an approach may be unsuitable for \ABCP{}. 
For example, consider the Markov model of Example \ref{ex:markovmodel} where the summary statistic $\bary_{\teta}$ is informative for the model parameter $c$ and the statistic $y_n$, the last observation, is informative for prediction. 
We formed a discrepancy $\Delta(\sss_y,\sss_z)=(\bary_{\teta}-\barz_{\teta})^2/\Var(\barz_{\teta}) + (y_n-z_n)^2/\Var(z_n)$ where the variances were estimated via pilot simulations. We then observed that this common approach lead to too low weight for the statistic $y_n$ due to its fairly high variability (see Section \ref{subsubsec:abcpcomp}) and consequently noticeable ABC error in the prediction. 
This also happened when e.g.~robust measures of variability were used and in the more realistic case of the continuous-time Markov model of Section \ref{subsec:lv}. 
This difficulty can be alleviated by manual readjustment but this is a cumbersome and opaque procedure in practice. 

The use of multiple discrepancies (see Section \ref{subsec:abc}) can be helpful for appropriately weighting the summary statistics in some ABC prediction situations. 
Suppose that a meaningful partition of the summary statistic exists so that $\sss(y)=(\bar{\sss}(y),\tl{\sss}(y))$ where $\bar{\sss}(y)$ denotes those statistics that are deemed informative for the parameter $\theta$, and $\tl{\sss}(y)$ those that are deemed informative for prediction (and ideally also interpretable). In our Markov example these would be $\bar{\sss}(y)=\bary_{\teta}$ and $\tl{\sss}(y)=y_n$. 
Then one could use the kernel
\begin{equation}
    \indic_{\|\bar{\sss}_y-\bar{\sss}_z\| \leq \bar{h}_t}
    \indic_{\|\tl{\sss}_y-\tl{\sss}_z\|' \leq\tl{h}_t}
    = \indic_{\sss_z\in\{u=(\bar{u},\tl{u})\in\reals^d : \|\bar{\sss}_y-\bar{u}\| \leq \bar{h}_t, \|\tl{\sss}_y-\tl{u}\|' \leq \tl{h}_t\}},
    \label{eq:twothresholds}
\end{equation}
where $\bar{h}_t$ and $\tl{h}_t$ denote separate thresholds and $\|\cdot\|$, $\|\cdot\|'$ the corresponding norms. 
If one chooses $\bar{h}_t/a=\tl{h}_t/b=h_t$ where $a,b>0$ are fixed constants and $h_t$ a sequence of thresholds such that $h_t\rightarrow 0$ as $t\rightarrow\infty$, then the acceptance region of (\ref{eq:twothresholds}) can be written as $\A_t=\{u=(\baru,\tl{u})\in\reals^d:\max\{a\|\bar{\sss}_y-\bar{u}\|,b\|\tl{\sss}_y-\tl{u}\|'\}\leq h_t\}$. It is easy to check that $\A_t$ satisfies the assumptions \ref*{it:a2}, \ref*{it:a4} and \ref*{it:a5} so that Proposition \ref{thm:abcp} applies. 
An important technical condition is that $\bar{h}_t$ and $\tl{h}_t$ decay at the same rate, otherwise the assumption \ref*{it:a5} would not hold. 
In our experiments in Section \ref{sec:experiments} where the above approach is used, $||\cdot||'$ and $\tl{h}_t$ are selected based on expert knowledge and the scale of the observed data, respectively. 
The weights for $\|\cdot\|$ can be determined using pilot runs or adaptively. 
The best practices likely depend on the problem at hand and a detailed investigation falls outside of the scope of this paper.

\section{Illustrative examples} \label{sec:experiments}

We next demonstrate the predictive ABC methods of Section \ref{sec:main} using two intractable dynamic models, M/G/1 queue (Section \ref{subsec:mg1}) and stochastic Lotka-Volterra (Section \ref{subsec:lv}). 
The goal of these numerical experiments is two-fold: First, to show that \ABCP{} and \ABCL{} can produce reasonable approximations of the posterior predictive distribution in practice, although overestimated uncertainty often results as is common also in the standard ABC inference. 
Second, the results indicate that the goal of prediction should be accounted for when designing the summary statistics, although informative low-dimensional summary statistics (both for parameter estimation and prediction) are challenging to design for realistic applications. 
While some insight on selecting informative summaries and suitable ABC sampler configurations for the goal of prediction is provided, we do not aim for a comprehensive analysis. This is because these choices depend on the problem at hand and are active topics of ongoing research also in the context of standard ABC inference. 

\ABCF{} and \ABCL{} approaches do not apply in all of our test cases unlike \ABCP{} which, in principle, can be implemented whenever jointly sampling from $\pdf(\tl{y},y\cond\theta)$ is feasible. 
Importantly, all three approaches produce the standard ABC posterior for $\theta$ as their marginal whenever they can be implemented. Hence, whenever we additionally computed this quantity, we only show it for \ABCP{} and the reader needs to notice that \ABCF{} and \ABCL{} based on the same summary statistic, discrepancy and threshold as \ABCP{} would also produce the same result. 
R-code (with fast C-implementation for model simulation) used for the experiments is available at \url{https://github.com/mjarvenpaa/ABC-pred-inf}.

\subsection{M/G/1 queue model} \label{subsec:mg1}

We first consider the M/G/1 queue model where customers arrive at a single server with independent interarrival times $\mgw_i\sim\Exp(\theta_3)$, that is, the arrival times follow Poisson process with rate parameter $\theta_3$. 
The queue is empty before the first arrival. 
The service times $\mgu_i$ are assumed to be independently distributed so that $\mgu_i\sim\Unif([\theta_1,\theta_2])$. Only the interdeparture times $\mgy=(\mgy_1,\ldots,\mgy_n)$ are observed. 
This model has been used before e.g.~by \citet{Fearnhead2012,Jiang2018,Bernton2019} to demonstrate ABC algorithms for estimating the model parameters $\theta=(\theta_1,\theta_2,\theta_3)$. We also assume that the parameters $\theta$ are unknown but the main task is to estimate the waiting times of future customers. 

The interdeparture times $\mgy_i$ satisfy
\begin{equation}
    \mgy_i = \mgu_i + \max\bigg\{ 0, \sum_{j=1}^i \mgw_j - \sum_{j=1}^{i-1} \mgy_j \bigg\}
    = \mgu_i + \max\left\{ 0, \mgv_i - \mgx_{i-1} \right\},
    \label{eq:mg1}
\end{equation}
where $\mgv_i\eqdef\sum_{j=1}^i \mgw_j$ are the arrival times and $\mgx_i\eqdef\sum_{j=1}^i \mgy_j$ the corresponding departure times. We use the convention $\mgv_0=\mgx_0=0$.
While the likelihood function $\pdf(\mgy\cond\theta)$ is intractable as discussed by \citet{Heggland2004}, using (\ref{eq:mg1}) and as $\mgv_i=\mgv_{i-1}+\mgw_i, \mgw_i \sim \Exp(\theta_3)$, one can easily simulate from 
\begin{equation}
    \pdf(\mgy_i,\mgv_i\cond\mgy_{1:i-1},\mgv_{1:i-1},\theta)
    = \pdf(\mgy_i,\mgv_i\cond\mgx_{i-1},\mgv_{i-1},\theta) 
    = \pdf(\mgy_i\cond\mgx_{i-1},\mgv_i,\theta_{1:2})\pdf(\mgv_i\cond\mgv_{i-1},\theta_3)
\end{equation}
and hence from $\pdf(\mgy,\mgv\cond\theta)$, where $\mgv=(\mgv_1,\ldots,\mgv_n)$. Obviously, this way one can sample also from 
$\pdf(\tl{\mgy},\mgy\cond\theta)$ where $\tl{\mgy}=(\mgy_{n+1},\ldots,\mgy_{n+\tl{n}})$ denote the interarrival times of the next $\tl{n}$ future customers. 

We use $\mgwait_i \eqdef \mgx_i-\mgv_i, i=1,\ldots,n$, to denote the waiting times of customers and similarly $\tl{\mgwait}_i, i = n+1,\ldots,n+\tl{n}$, denote those of the future customers. This definition includes both the time the customer waits to be served (which can be $0$) and the service time.
Clearly, sampling from $\pdf(\tl{\mgwait},\mgy\cond\theta)$ is feasible so \ABCP{} can be easily implemented. Similarly, simulating from $\pdf(\tl{\mgwait}\cond \mgy,\mgv,\theta) = \pdf(\tl{\mgwait}\cond \mgx_n,\mgv_n,\theta)$ is feasible which facilitates \ABCL{}. 
Simulating from $\pdf(\tl{\mgy}\cond \mgy, \theta)$ or $\pdf(\tl{\mgwait}\cond \mgy, \theta)$, and consequently implementing \ABCF{}, is not straightforward due to the latent variables $v$ involved and is hence not considered\footnote{Since the model is neither black-box nor highly complex, it might be possible to devise an algorithm to sample from $\pdf(\tl{\mgwait}\cond \mgy, \theta)$. Whether such algorithm could be used for this particular case is however irrelevant for our aims here.}.

\subsubsection{Common experimental details} \label{subsubsec:mg1details}

We use the uniform prior $(\theta_1,\theta_2-\theta_1,\theta_3) \sim \Unif([0,10]^2\times[0,1/3])$. 
In addition, we account for the constraint {\color{\revcol}$\theta_1 \leq \min_{i\in\{1,\ldots,n\}}\mgy_i$}, which is is implicitly coded into the intractable likelihood function but not in the ABC procedure, as discussed by \citet{Bernton2019}. 
Although M/G/1 model has commonly been used to study ABC algorithms, particle MCMC methods \citep{Andrieu2010} and the special MCMC algorithm by \citet{Shestopaloff2014} in principle also apply. 
We adopt the latter MCMC method and we further extend it in a straightforward way to obtain samples from $\pdf(\tl{\mgwait}\cond y)$. This allows us to compare the accuracy of the ABC methods to the ground-truth. 
We also consider the predictive distribution $\pdf(\tl{\mgwait}\cond \mgy, \mgv, \thetatrue)$ as this unveils the best possible prediction accuracy. Of course, in reality this baseline would be unavailable as $\thetatrue$ is unknown and $v$ is unobserved. 

\begin{figure}[htbp] 
\centering
\begin{subfigure}[b]{0.31\textwidth}
\centering
\includegraphics[width=\textwidth]{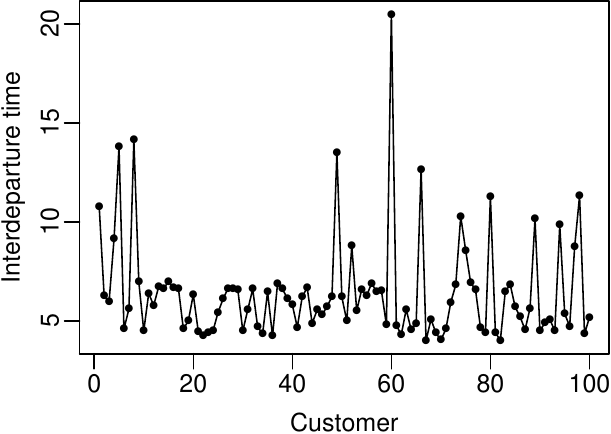}
\end{subfigure}
\hspace{0.3cm}
\begin{subfigure}[b]{0.31\textwidth}
\centering
\includegraphics[width=\textwidth]{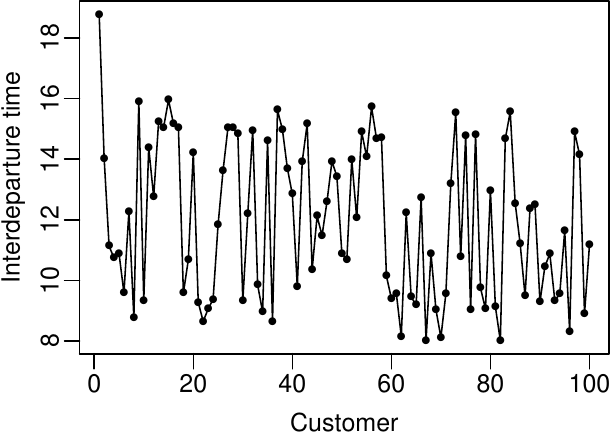}
\end{subfigure}
\caption{Observed interdeparture times $\mgy=(\mgy_1,\ldots,\mgy_n)$ with $n=100$ used in the M/G/1 experiments. \emph{Left plot:} The case of Section \ref{subsec:mg1case1} where the queue length is varying. Occasional large interdeparture times suggest that the queue could be then empty. \emph{Right plot:} The case of Section \ref{subsec:mg1case2} where the queue length tends to be growing.} \label{fig:mg1datasets}
\end{figure}

We form a common $5$-dimensional baseline summary statistic $\sss^{(0)}$ by using the $\alpha=0.25$, $0.5$ and $0.75$th empirical quantiles $\hat{q}_{\alpha}(\mgy)$, and the range of $\mgy$. 
For comparison, we adopt the strategy outlined in Section \ref{subsec:formingdiscr} and form another summary statistic $\sss^{(1)}(\mgy) = (\bar{\sss}^{(1)}(\mgy),\tl{\sss}^{(1)}(\mgy))$, where $\bar{\sss}^{(1)}(\mgy) = {\sss}^{(0)}(\mgy)$ and where the additional summaries $\tl{\sss}^{(1)}(\mgy)$ are given by $\max\{i\geq 1 : \mgy_i\geq \hat{q}_{\tl{\alpha}}(\mgy_{\textnormal{obs}})\}$ with convention $\max\varnothing=1$. Note that $\mgy$ denotes any data realization while $\hat{q}_{\tl{\alpha}}(\mgy_{\textnormal{obs}})$ is computed using the observed data at hand denoted here as $\mgy_{\textnormal{obs}}$ for clarity. This tentative summary statistic captures the idea that individual large interdeparture times result when the queue is empty and the locations of such recent values provide information about the state of the queue for prediction. 
We use this strategy and $\sss^{(1)}$ also for \ABCL{}. 
The statistic $\tl{\sss}^{(1)}(\mgy) = \mgy_n$ is not chosen as it would be rather uninformative -- short interdeparture times occur frequently irrespective of the state of the queue as seen in Figure \ref{fig:mg1datasets}.

We use $\tl{\alpha}=$ 0.7, 0.8 and 0.9 so that $\tl{\sss}^{(1)}(\mgy)\in\reals^3$. 
We choose $||\cdot||$ for both $\sss^{(0)}$ and $\bar{\sss}^{(1)}$ so that $||r|| = r\T \hat{C}^{-1}r$ where $\hat{C}$ is the empirical covariance matrix of the summaries estimated, for simplicity and to facilitate meaningful comparison, by running the model $10^3$ times at $\thetatrue$. This value would be unavailable in practice but other point estimates or the approach by \citet{Prangle2017} could be then used instead. 
We choose $||\cdot||'$ as the $L^{\infty}$-norm so that $||r||' = \max_{i\in\{1,2,3\}}|r_i|$. 
%
We use ABC-MCMC with $2\cdot 10^6$ iterations and burn-in $10^4$ in both cases. In the case of $\sss^{(0)}$, the threshold $h$ is determined so that the acceptance probability (after burn-in) is $2\%$. In the case of $\sss^{(1)}$ we set $\tl{h}=5$ (Section \ref{subsec:mg1case1}) or $\tl{h}=7$ (Section \ref{subsec:mg1case2}). The threshold $\bar{h}$ is then determined to give the same acceptance rate as with $\sss^{(0)}$ to make the comparison meaningful although it is not possible to completely separate the effect of the summary statistic and choices related to computational efficiency.

\subsubsection{Varying queue} \label{subsec:mg1case1}

We first consider a situation where $\thetatrue=(4, 7, 0.15)$. In this case the queue length tends to vary being occasionally empty and occasionally containing several customers. A simulated data set with $n=100$ used in the experiments below is shown in Figure~\ref{fig:mg1datasets}. 

\begin{figure}[htb] 
\centering
\begin{subfigure}{0.7\textwidth}
\centering
\includegraphics[width=\textwidth]{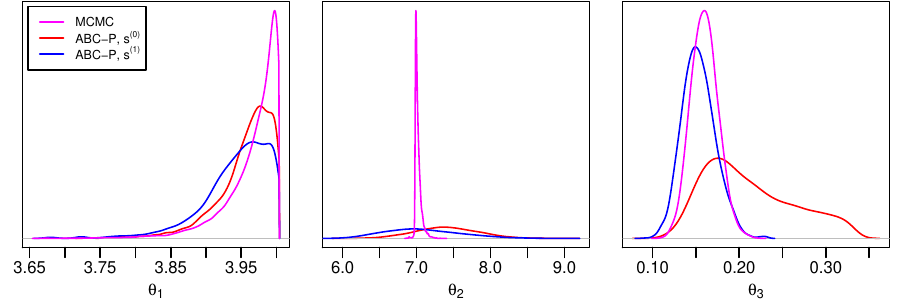}
\end{subfigure}
\\[0.2cm]
\centering
\begin{subfigure}{0.7\textwidth}
\centering
\includegraphics[width=\textwidth]{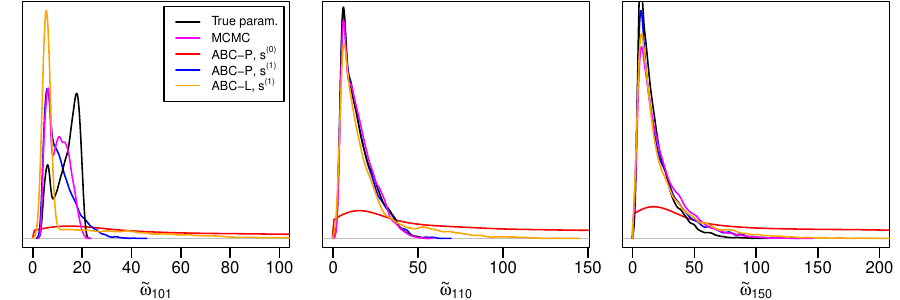}
\end{subfigure}
\caption{Results for the M/G/1 experiments of Section \ref{subsec:mg1case1}. \emph{Top row:} Marginal posterior distributions computed using MCMC and ABC for parameters $\theta$. 
\emph{Bottom row:} Predictive distribution based on the true parameter $\thetatrue$ (``True param.'') and posterior predictive distributions computed using MCMC and ABC approaches for the waiting times of some future customers. The right tail of the \ABCP{} posterior predictive based on $\sss^{(0)}$ is truncated to ease visualization.} \label{fig:mgc1}
\end{figure}

The top row of Figure \ref{fig:mgc1} shows the ABC posteriors of the parameter $\theta$. 
The marginal ABC posteriors for $\theta_1$ are both fairly accurate, though the baseline summary statistic $\sss^{(0)}$ works slightly better. The likely reason is that the threshold $\bar{h}$ was set slightly larger than $h$ because the additional summaries $\tl{\sss}^{(1)}$ need also to be matched in this case and because the ABC-MCMC acceptance probability was required to be the same in both cases. 
On the other hand, $\sss^{(1)}$ works better for $\theta_3$ which suggests that the additional summaries $\tl{\sss}^{(1)}$ designed for prediction are informative also for this parameter. As common also in other ABC analyses in literature, poor marginal posterior approximations for $\theta_2$ are observed. 

The bottom row of Figure \ref{fig:mgc1} shows the corresponding results for the prediction task. Interestingly, the ideal predictive density and the exact posterior predictive for the first future waiting time $\mgwait_{101}$ are both bimodal. (We run several MCMC chains to ensure our observations are not caused by possible numerical issues.) 
An important result is that the \ABCP{} posterior predictive based on the baseline summary statistic $\sss^{(0)}$ has far too long right tail. This occurs because $\sss^{(0)}$ is invariant to the order of the observations and hence the trend of the observed interdeparture times just before the prediction gets lost. 
This would also be the outcome with other methods in literature such as the distance functions studied by \citet{Jiang2018,Bernton2019}. 
\ABCL{} posterior predictive also has a pronounced right tail but the additional summaries $\tl{\sss}^{(1)}$ appear informative for prediction so that \ABCP{} based on $\sss^{(1)}$ produces reasonable approximations.

\subsubsection{Growing queue} \label{subsec:mg1case2}

We use $\thetatrue=(8, 16, 0.15)$ in our second example so that the arrivals are relatively frequent as compared to the service times. 
The queue is mostly full in this case and the interarrival times $\mgy_i$ have approximately the same distribution as the service times $\mgu_i$. We observe that $\mgwait_i-\mgwait_{i-1} = \mgy_i-\mgw_i$
then approximately follows the same distribution as $\mgu_i-\mgw_i$. That is, the increments of the process $\{\mgwait_i\}_{i\geq 1}$ tend to be \iid{}~distributed so that $\{\mgwait_i\}_{i\geq 1}$ is approximately a random walk. Furthermore, as $\mgu_i-\mgw_i$ tends to be positive, the waiting times approximately grow linearly in expectation. 
The data set used is shown in Figure~\ref{fig:mg1datasets}. 

\begin{figure}[htbp] 
\centering
\begin{subfigure}{0.7\textwidth}
\centering
\includegraphics[width=\textwidth]{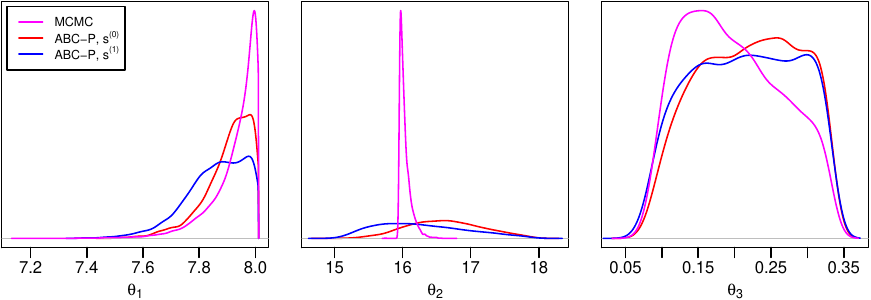}
\end{subfigure}
\begin{subfigure}{0.28\textwidth}
\centering
\includegraphics[width=\textwidth]{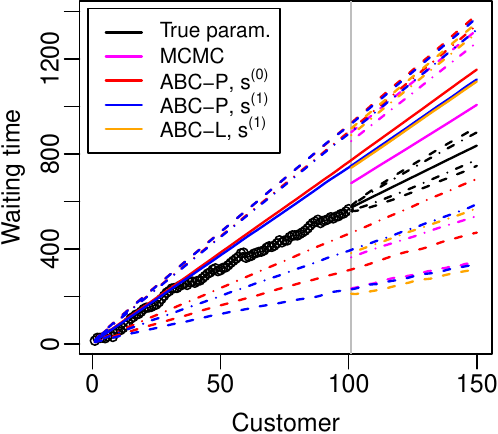}
\end{subfigure}
\caption{Typical results for the M/G/1 experiment of Section  \ref{subsec:mg1case2}. \emph{The first three plots} on the left show the marginal posteriors for $\theta$. \emph{The fourth plot} illustrates the predictive densities for $\mgwait$ (solid lines: the median, dot-dashed lines: 75\% credible interval (CI), dashed lines: 90\% CI). Note that these lines are also drawn for the observed customers in the case of \ABCP{}. The vertical gray line shows the first future customer. The black rings show the unobserved waiting times not used for inference.} \label{fig:mgc2}
\end{figure}

Figure \ref{fig:mgc2} shows that the marginal ABC posterior distributions of $\theta_1$ and $\theta_2$ feature similar trends as before while those of $\theta_3$ are equally good this time. The posterior uncertainty of $\theta_3$ is however noticeable because the observed interdeparture times are essentially determined by the service times that depend only on $\theta_1$ and $\theta_2$. 
Similarly, the observations are not very informative for estimating future waiting times and the ideal predictive density substantially benefits from its use of the true values of $\theta$ and $\mgv$. Figure \ref{fig:mgc2} further shows that the waiting times are growing approximately linearly as the above analysis suggests. 
All ABC approaches produce similar predictive densities which are comparable to the ground-truth. It is intuitive that the knowledge of the order of the interdeparture times or the additional summaries in $\sss^{(1)}$ are not highly useful when the queue is growing fairly steadily. 

\subsection{Stochastic Lotka-Volterra model} \label{subsec:lv}

We consider the Lotka–Volterra model which is a Markov jump process that describes the continuous-time evolution of a population of predators interacting with a prey population. This model also serves as a basic example of stochastic kinetics networks used to model the evolution of interacting molecules, see e.g.~\citet{Wilkinson2018}. 
The size of the prey population at time $t$ is $\yone_t$ and similarly $\ytwo_t$ is the predator population. We also denote $y_t \eqdef (\yone_t,\ytwo_t)$. The model dynamics are Markovian with the transition probabilities
\begin{equation}
    \prob(y_{t+\deltat}=(w_1',w_2') \cond y_{t}=(w_1,w_2), \theta) 
    = \begin{cases}
    1-\gamma_{\theta}(w)\deltat + \smallo(\deltat), & \textnormal{if } w_1'=w_1 \textnormal{ and } w_2'=w_2, \\
    \theta_1 w_1\deltat + \smallo(\deltat), & \textnormal{if } w_1'=w_1+1 \textnormal{ and } w_2'=w_2, \\
    \theta_2 w_1 w_2\deltat + \smallo(\deltat), & \textnormal{if } w_1'=w_1-1 \textnormal{ and } w_2'=w_2+1, \\
    \theta_3 w_2\deltat + \smallo(\deltat), & \textnormal{if } w_1'=w_1 \textnormal{ and } w_2'=w_2-1, \\
    \smallo(\deltat), & \textnormal{otherwise},
    \end{cases}
    \label{eq:lvdyn}
\end{equation}
where $\gamma_{\theta}(w) \eqdef \theta_1 w_1 + \theta_2 w_1 w_2 + \theta_3 w_2$, $\deltat$ is a small time increment and $\theta = (\theta_1,\theta_2,\theta_3)$ are positive rate parameters. 

While the likelihood function associated with this model is intractable (except for some special scenarios), realizations can be simulated given any initial state and parameters $\theta$ using the Gillespie algorithm \citep{Gillespie1977}. 
The Markov property implies that one can also simulate from $\pdf(\tl{y}\cond y,\theta) = \pdf(\tl{y}\cond y_{\tp},\theta)$ where $\tl{y}$ represent the population sizes at some future time points, $y$ the observed populations and $y_{\tp}$ the observed populations at the last observation time $\tp$. 
However, if e.g.~only the prey population is observed then \ABCF{} approach cannot be directly implemented via Gillespie algorithm because $y_{\tp}$ is not fully known. 
Particle MCMC methods can be in principle used for inference in the noisy case, see \citet{Wilkinson2018,Warne2020} and references therein for details. In this paper we however limit our attention to ABC methods in the non-noisy setting.

\subsubsection{Common experimental details} \label{subsubsec:lvdetails}


The following settings are common to all Lotka-Volterra experiments considered: 
The initial state is $y_0=(100,50)$ and we use $\thetatrue=(1, 0.005, 0.6)$ which produces oscillating behavior of the populations. The parameters are log-transformed for inference and we use the prior $(\log\theta_1,\log\theta_2,\log\theta_3) \sim \Unif([-6,2]^3)$. 
%
We use similar strategies for forming the summary statistics, selecting thresholds and the same settings for the ABC-MCMC sampler as in the M/G/1 experiments. 
{\color{\revcol}In particular, the acceptance probability of ABC-MCMC is $2\%$ and $2\cdot 10^6$ iterations are run}. We use $L^{\infty}$-norm for $||\cdot||'$ but this time $||\cdot||$ is the weighted Euclidean distance whose weights are reciprocals of the squared median absolute deviations (MAD) of the simulated summaries estimated empirically as before. 
Summary statistics are designed individually for each inference task as explained below.

\subsubsection{Prediction tasks} \label{subsec:lvpred}

We consider a prediction task where the population sizes at $81$ equidistant times in $[0,\ta]$ are observed and the main goal is to compute the posterior predictive distribution of both populations in $[\ta,\tb]$. We select $\ta=24$ and $\tb=45$ time units. 
We study two cases: 1) Both populations are observed and we compare \ABCP{} and \ABCF{} and 2) Only prey populations $\yone_t$ are observed and we compare \ABCP{} and \ABCL{}. \ABCL{} is implemented by setting $v=\ytwo_{\ta}$ in case 2. 

Similarly to earlier analyses (e.g.~\citet{Papamakarios2016,Wilkinson2018}) where the goal is to infer $\theta$, we use the following summary statistics for our case 1: the mean, standard deviation, two first autocorrelations of both populations and cross-correlation between both populations. The resulting $9$-dimensional baseline summary statistic is denoted by $\sss^{(0)}$. 
We then form a summary statistic
$\sss^{(1)}$ specifically for the prediction task at hand: We set $\bar{\sss}^{(1)} = \sss^{(0)}$ and $\tl{\sss}(y) = y_{\ta}\in\reals^2$. 
%
{\color{\revcol}
The baseline summary statistic $\sss^{(0)}$ for case 2 includes the mean, standard deviation and two first autocorrelations of the prey population. 
Summary statistic $\sss^{(1)}$ is formed so that $\bar{\sss}^{(1)} = \sss^{(0)}$ and\footnote{{\color{\revcol}While $y_{\ta}$ is sufficient in the sense of (\ref{eq:predsuffstrong2}) in case 1 due to the Markov property, this does not hold for $\yone_{\ta}$ when only the prey is observed. Due to this fact and guided by preliminary experiments we include additional observations to $\tl{s}$.}} $\tl{\sss}(y) = (\yone_{\ta-i\deltameas})_{i=0}^{5}\in\reals^{6}$ where $\deltameas=0.3$ is the time between consecutive observations. 
We use $\tl{h}=50$ in case 1 and $\tl{h}=30$ in case 2.}
{\color{\revcol}In case 1 we compare our ABC methods to the ideal predictive distribution $\pdf(\tl{y}\cond y, \thetatrue)=\pdf(\tl{y}\cond y_{\ta}, \thetatrue)$ from which we directly simulate using the Gillespie algorithm. In case 2 our comparison is to an approximation of $\pdf(\tl{y}\cond \yone, \thetatrue)$ which is obtained by first generating $(\tl{y}^{(i)},z^{(i)}) \sim \pdf(\tl{y},z\cond \thetatrue), i=1,\ldots,10^6$, using the Gillespie algorithm and accepting $\tl{y}^{(i)}$ if $||\tl{\sss}(z^{(i)})-\tl{\sss}(y)||'=\max_{i\in\{0,\ldots,5\}}|\zonei_{\ta-i\deltameas}-\yone_{\ta-i\deltameas}|\leq 7$. This is a form of ABC rejection sampling (but conditionally on $\thetatrue$, see \appe{} \ref{appe:abcpalt} for discussion).}

\begin{figure}[htbp] 
\centering
\begin{subfigure}[b]{0.48\textwidth}
\centering
\includegraphics[width=\textwidth]{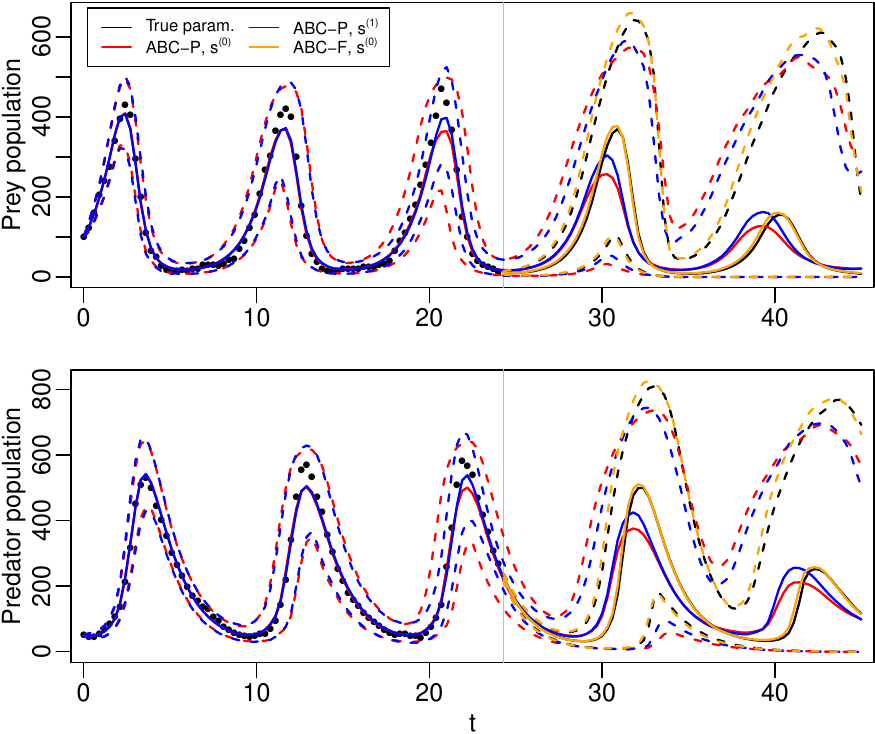}
\caption{Case~1: Both populations observed.}
\label{fig:lvpred1a}
\end{subfigure}
\hfill
\begin{subfigure}[b]{0.48\textwidth}
\centering
\includegraphics[width=\textwidth]{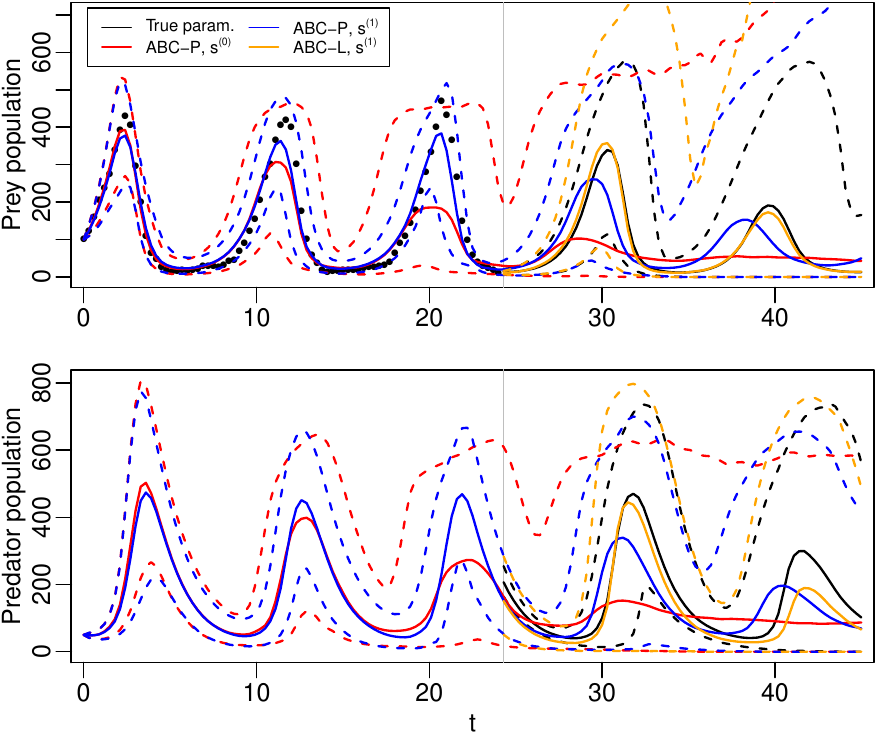}
\caption{Case~2: Only prey population observed.}
\label{fig:lvpred1b}
\end{subfigure}
\caption{Typical results for the Lotka-Volterra experiments of Section \ref{subsec:lvpred}. The vertical gray line shows where the prediction interval starts and the black circles show the observations. The solid lines show the median and the dashed lines the $90\%$ CI. These lines are also drawn in $[0,\ta]$ to demonstrate that the observed data is not conditioned on exactly in ABC. ``True param.''~shows the ideal predictive distribution based on $\thetatrue$ {\color{\revcol}(case 1) and its approximation (case 2)}. The prey population size is truncated in case 2 because the predator population can die out with a non-negligible posterior probability in which case the prey population starts to grow exponentially fast.} \label{fig:lvpred1}
\end{figure}

Typical results for case 1 are shown in Figure~\ref{fig:lvpred1a}. 
{\color{\revcol}All ABC posterior predictive densities roughly capture the future model dynamics. 
The mean absolute error of the posterior median prediction (abbreviated as MAE and where the mean is taken over both populations and the prediction interval) of \ABCP{} is $40$ with both $\sss^{(0)}$ and $\sss^{(1)}$. Statistic $\sss^{(1)}$ however produces smaller standard deviation (std) of the posterior predictive in the region $t\approx\ta$ than $\sss^{(0)}$ (e.g.~$27$ vs.~$70$ for the predator population at $t=\ta+\deltameas$). 
While both approaches produce visibly inflated $90\%$ CI there, the accuracy with $\sss^{(1)}$ could be increased at the cost of more simulations by lowering $\tl{h}$ as this quantity directly controls the absolute error at $t=\ta$.} However, we do not have such apparent guarantee with $\sss^{(0)}$. 
The \ABCF{} posterior predictive is similar to the ideal predictive density over the whole interval $[\ta,\tb]$ {\color{\revcol}(MAE is $5.3$ and the std of the posterior predictive of the predator population at $t=\ta+\deltameas$ is $6.6$ while that of the ideal baseline is $6.4$)} so that \ABCF{} is clearly preferable to \ABCP{}. 

\begin{figure}[htbp] 
\centering
\begin{subfigure}{0.7\textwidth}
\centering
\includegraphics[width=\textwidth]{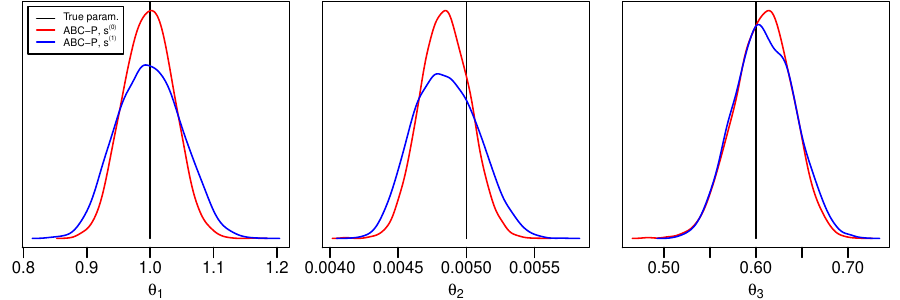}
\end{subfigure}
\\[0.2cm]
\begin{subfigure}{0.7\textwidth}
\centering
\includegraphics[width=\textwidth]{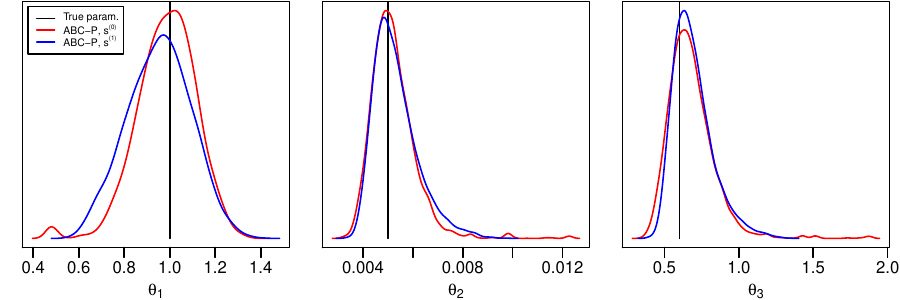}
\end{subfigure}
\caption{Posterior distributions for the parameters of the Lotka-Volterra experiments corresponding to Figure~\ref{fig:lvpred1} and Section \ref{subsec:lvpred}. \emph{Top row:} Case 1 where both populations are observed. \emph{Bottom row:} Case 2 where only prey population is observed. {\color{\revcol}The black vertical line shows the true value of the parameter.}} \label{fig:lvpred2}
\end{figure}

Figure~\ref{fig:lvpred1b} shows that, unsurprisingly, prediction in case 2 is more challenging than in case 1.
{\color{\revcol}
\ABCP{} based on $\sss^{(1)}$ produces visibly more accurate posterior predictive than $\sss^{(0)}$. MAE for \ABCP{} with $\sss^{(0)}$ is $72$ and with $\sss^{(1)}$ it is $54$.} 
{\color{\revcol}\ABCL{} produces the most accurate median prediction (MAE is $24$) and also the least inflated posterior predictive of the prey population in the region $t\approx\ta$ as a consequence of its ability to partially use the exact value of the observation $\yone_{\ta}$. \ABCP{} based on $\sss^{(1)}$ and  \ABCL{} however both produce similar predictive posteriors for the unobserved predator population $\ytwo_{t}$ at $t\approx\ta$.}
Marginal ABC posterior distributions for $\theta$ in both cases are finally shown Figure \ref{fig:lvpred2}. 
{\color{\revcol}Summary statistic $\sss^{(1)}$ produces slightly more inflated posterior as $\sss^{(0)}$ similarly -- and for the same reason -- as in the M/G/1 experiments.}

\subsubsection{Missing data task} \label{subsec:lvmis}

The goal here is to estimate the unobserved population sizes in $[\ta,\tb]$ based on $51$ equidistant observations both in $[0,\ta]$ and $[\tb,\tc]$ with $0<\ta<\tb<\tc$. In particular, we use $\ta=15, \tb=36$ and $\tc=51$. 
The summary statistics of Section \ref{subsec:lvpred} are not meaningful and we instead use the mean and standard deviation of both population sizes separately computed for intervals $[0,\ta]$ and $[\tb,\tc]$. This produces $8$-dimensional summary statistic $\bar{\sss}^{(1)}$. Additionally, we choose $\tl{\sss}^{(1)}(y) = (y_{\ta},y_{\tb})\in\reals^4$ and $\tl{h}=50$. 
Only \ABCP{} is considered as the other ABC approaches do not to apply\footnote{{\color{\revcol}Special methods called bridges have been proposed to approximately draw from $\pdf(\tl{y}\cond y_{\ta}, y_{\tb}, \theta)$ in the case of some Markov jump processes, see \citet{Golightly2019}. It seems possible to combine them with \ABCF{} but we do not analyze the feasibility of such model-specific approach in this paper.}}. 
We compare it to the ideal predictive distribution {\color{\revcol}$\pdf(\tl{y}\cond y, \thetatrue)=\pdf(\tl{y}\cond y_{\ta}, y_{\tb}, \thetatrue)$ which is computed in a similar fashion as the one in case 2 of Section \ref{subsec:lvpred}: We first generate $10^6$ samples $(\tl{y}^{(i)},z_{\tb}^{\smash{(i)}}) \sim \pdf(\tl{y},z_{\tb}\cond y_{\ta}, \thetatrue)$ and then accept those of them that satisfy $||z_{\tb}^{\smash{(i)}}-y_{\tb}||'\leq 5$ (our first data set) or that match the observed $y_{\tb}$ exactly (our second data set). In the latter case we obtain exact samples from $\pdf(\tl{y}\cond y_{\ta}, y_{\tb}, \thetatrue)$.}

\begin{figure}[htbp]
\centering
\begin{subfigure}[b]{0.48\textwidth}
\centering
\includegraphics[width=\textwidth]{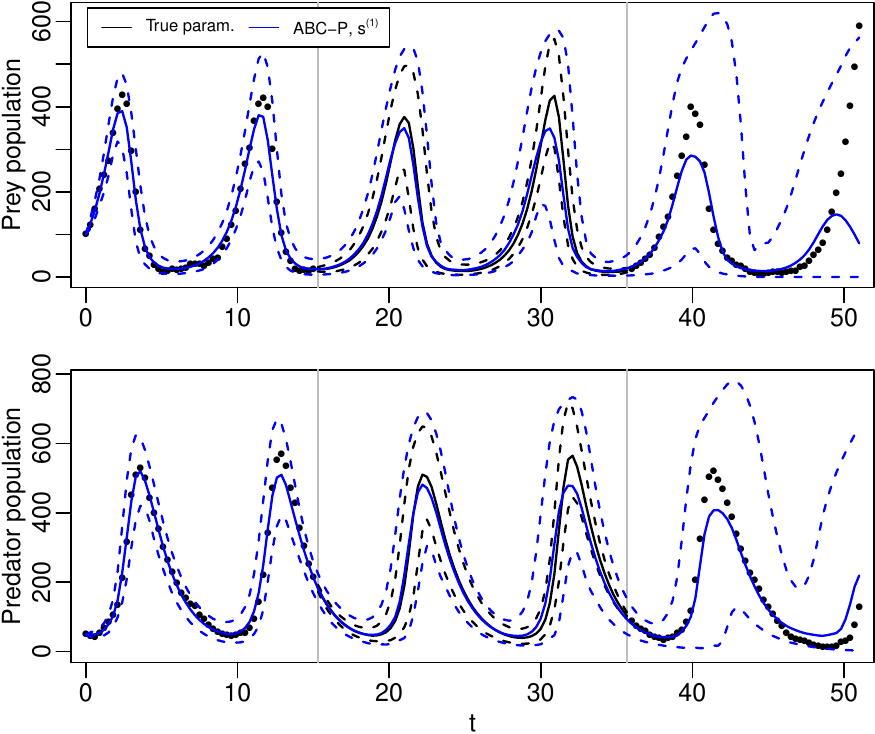}
\end{subfigure}
\hfill
\begin{subfigure}[b]{0.48\textwidth}
\centering
\includegraphics[width=\textwidth]{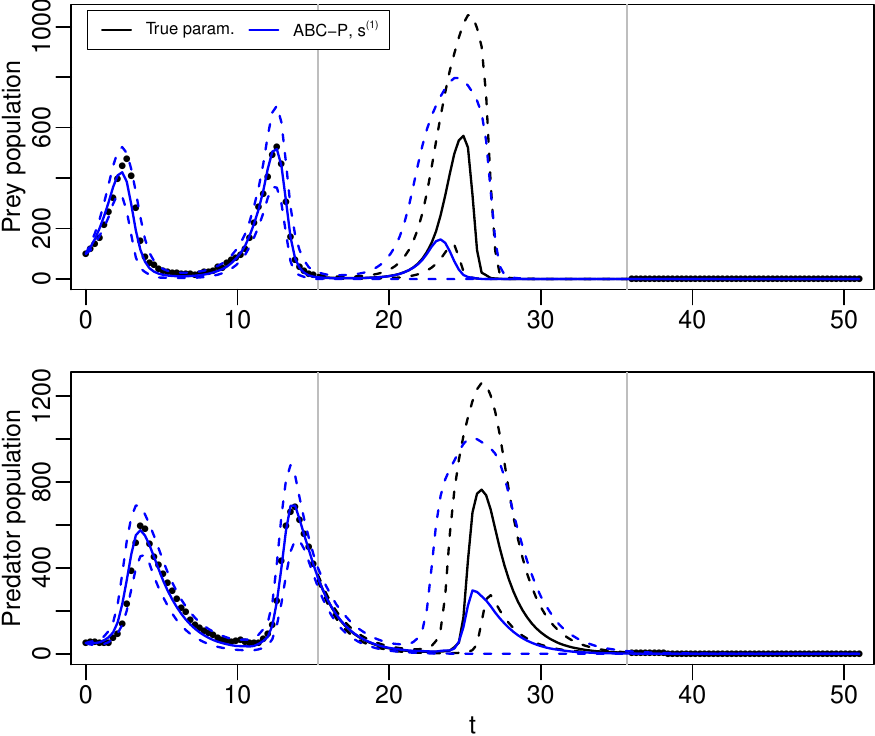}
\end{subfigure}
\caption{Results for the Lotka-Volterra experiment of Section \ref{subsec:lvmis} with two sets of observed data. {\color{\revcol}The vertical gray lines indicate the unobserved time interval $[15,36]$ and the black circles show the observations. The solid lines show the median and the dashed lines the $90\%$ CI. ``True param.''~shows the ideal predictive distribution based on $\thetatrue$ (right column) and its approximation (left column).}} \label{fig:lvmis1}
\end{figure}

Results with two data realizations are shown in Figure~\ref{fig:lvmis1}. In a typical case shown in the left column, the approximation is good as compared to the approximate ideal predictive density {\color{\revcol}(MAE is $17$)}. As before, the uncertainty is overestimated especially near $\ta$ and $\tb$ as the matching to the observed $y_{\ta}$ and $y_{\tb}$ is only done up to the threshold $\tl{h}$. The right column presents a special situation where both populations have become extinct. The approximation quality is fair in this case {\color{\revcol}(MAE is $59$)}. 
{\color{\revcol}The ABC posteriors for $\theta$ are additionally shown in \appe{} \ref{appe:experiments}.}

\section{Discussion} \label{sec:conclusions}

There is no single natural, general way to to approximate the posterior predictive distribution of intractable models via ABC. In this paper we have studied three such complementary approaches summarized in Table \ref{table:abcpredoverview}. These methods make different assumptions regarding which predictive density of the model can be simulated from. 
We provided some new insight to the earlier approach, abbreviated as \ABCF{} in this paper, and pointed out its limitations. 
We then proposed related but more general \ABCP{} and \ABCL{} methods, investigated their basic properties (especially the selection of summary statistics), discussed some ways to implement these methods in practice, and finally provided numerical examples of all three methods. 

Our analysis indicates that forming summary statistics for \ABCP{} and weighting them appropriately requires more care than for the standard ABC inference or for \ABCF{} because these statistics need to be informative in the sense of predictive sufficiency instead of ordinary sufficiency. Of course, they should also be low-dimensional to avoid the curse of dimensionality. 
Such suitable statistics may have large variability which can lead to more pronounced computational challenges than in standard ABC inference and in some situations result substantially inflated predictive intervals. Also, suitable low-dimensional statistics may not exist which could be the case e.g.~with intractable spatial data models which were however not analyzed in this paper. 
On the other hand, such suitable statistics can be often constructed when the model features Markovian or some other simplifying structure. 
\ABCP{} produced reasonable posterior predictive approximations in our experiments although we used a basic ABC-MCMC implementation and summary statistics designed for prediction in a rather rudimentary fashion.

Our analysis and empirical results suggest that in general \ABCF{} should be used instead of \ABCP{}. 
However, as we have discussed, implementing \ABCF{} can be difficult or infeasible which makes \ABCP{} and \ABCL{} a useful addition to statistician's toolbox. 
We cannot conclude whether \ABCL{} should be preferred to \ABCP{} when both approaches are feasible. This likely depends on the model at hand and the suitability of the selected summary statistics which is often hard to assess in practice. 
{\color{\revcol}It should also be noted that when asymptotically exact (particle) MCMC or other model-specific methods can be used, they are likely more accurate than the more generic, approximate methods analyzed in this paper.} 

Our key aim was to give a unified overview on approximating the posterior predictive via ABC in an accessible manner. 
As future work, one could study more closely how to best use such methods for specific models and inference tasks. 
One could also investigate if LFI techniques such as regression adjustment \citep[]{Beaumont2002,Blum2010}, Bayesian synthetic likelihood \citep{Price2018}, {\color{\revcol}ratio density estimation \citep{Thomas2022}} or conditional density estimation methods based on neural networks or Gaussian processes \citep{Papamakarios2016,Papamakarios2018,Grazian2019,Jarvenpaa2020_babc} can be used to improve the accuracy or computational efficiency of the \ABCP{} approach. 
Another important research direction would be to study how to construct informative summary statistics for predictive ABC inference in an automatic fashion. 
Such methods have been developed for standard ABC inference (see e.g.~\citet[Chapter~5]{Sisson2019}) and some of them could possibly be extended to the predictive setting. 
Also, our analysis on the summary statistics for prediction might offer new perspective for selecting them for standard ABC inference.

\subsection*{Acknowledgments}

This research was funded by the Norwegian Research Council FRIPRO grant no.~299941 and by the European Research Council grant no.~742158.

\bibliography{bib/refs_abcpred_v2}

\begin{thebibliography}{}

\bibitem[Andrieu et~al., 2010]{Andrieu2010}
Andrieu, C., Doucet, A., and Holenstein, R. (2010).
\newblock {Particle Markov chain Monte Carlo methods}.
\newblock {\em Journal of the Royal Statistical Society: Series B (Statistical
  Methodology)}, 72(3):269--342.

\bibitem[Barber et~al., 2015]{Barber2015}
Barber, S., Voss, J., and Webster, M. (2015).
\newblock {The rate of convergence for approximate Bayesian computation}.
\newblock {\em Electronic Journal of Statistics}, 9(1):80--105.

\bibitem[Beaumont et~al., 2009]{Beaumont2009}
Beaumont, M.~A., Cornuet, J.-M., Marin, J.-M., and Robert, C.~P. (2009).
\newblock {Adaptive approximate Bayesian computation}.
\newblock {\em Biometrika}, 96(4):983--990.

\bibitem[Beaumont et~al., 2002]{Beaumont2002}
Beaumont, M.~A., Zhang, W., and Balding, D.~J. (2002).
\newblock {Approximate Bayesian computation in population genetics}.
\newblock {\em Genetics}, 162(4):2025--2035.

\bibitem[Bernardo and Smith, 1994]{Bernardo1994}
Bernardo, J.~M. and Smith, A. F.~M. (1994).
\newblock {\em Bayesian Theory}.
\newblock John Wiley \& Sons.

\bibitem[Bernton et~al., 2019]{Bernton2019}
Bernton, E., Jacob, P.~E., Gerber, M., and Robert, C.~P. (2019).
\newblock Approximate {B}ayesian computation with the {W}asserstein distance.
\newblock {\em Journal of the Royal Statistical Society: Series B (Statistical
  Methodology)}, 81(2):235--269.

\bibitem[Biau et~al., 2015]{Biau2015}
Biau, G., Cérou, F., and Guyader, A. (2015).
\newblock {New insights into Approximate Bayesian Computation}.
\newblock {\em Annales de l'Institut Henri Poincaré, Probabilités et
  Statistiques}, 51(1):376--403.

\bibitem[Bj\o{}rnstad, 1996]{Bjornstad1996}
Bj\o{}rnstad, J.~F. (1996).
\newblock On the generalization of the likelihood function and the likelihood
  principle.
\newblock {\em Journal of the American Statistical Association},
  91(434):791--806.

\bibitem[Blum, 2010]{Blum2010}
Blum, M. G.~B. (2010).
\newblock {Approximate Bayesian Computation: a nonparametric perspective}.
\newblock {\em Journal of American Statistical Association},
  105(491):1178--1187.

\bibitem[Buckwar et~al., 2020]{Buckwar2020}
Buckwar, E., Tamborrino, M., and Tubikanec, I. (2020).
\newblock {Spectral Density-Based and Measure-Preserving ABC for Partially
  Observed Diffusion Processes. An Illustration on Hamiltonian SDEs}.
\newblock {\em Statistics and Computing}, 30(3):627–648.

\bibitem[Bürkner et~al., 2020]{Burkner2020}
Bürkner, P.-C., Gabry, J., and Vehtari, A. (2020).
\newblock Approximate leave-future-out cross-validation for {Bayesian} time
  series models.
\newblock {\em Journal of Statistical Computation and Simulation},
  90(14):2499--2523.

\bibitem[Calvet and Czellar, 2014]{Calvet2014}
Calvet, L.~E. and Czellar, V. (2014).
\newblock {Accurate Methods for Approximate Bayesian Computation Filtering}.
\newblock {\em Journal of Financial Econometrics}, 13(4):798--838.

\bibitem[Canale and Ruggiero, 2016]{Canale2016}
Canale, A. and Ruggiero, M. (2016).
\newblock {Bayesian nonparametric forecasting of monotonic functional time
  series}.
\newblock {\em Electronic Journal of Statistics}, 10(2):3265--3286.

\bibitem[Del~Moral et~al., 2012]{Moral2012}
Del~Moral, P., Doucet, A., and Jasra, A. (2012).
\newblock {An adaptive sequential Monte Carlo method for approximate Bayesian
  computation}.
\newblock {\em Statistics and Computing}, 22(5):1009--1020.

\bibitem[Drovandi et~al., 2022]{Drovandi2022}
Drovandi, C., Nott, D.~J., and Frazier, D.~T. (2022).
\newblock Improving the accuracy of marginal approximations in likelihood-free
  inference via localisation.
\newblock Available at \url{https://arxiv.org/abs/2207.06655}.

\bibitem[Fasiolo et~al., 2016]{Fasiolo2016}
Fasiolo, M., Pya, N., and Wood, S.~N. (2016).
\newblock {A Comparison of Inferential Methods for Highly Nonlinear State Space
  Models in Ecology and Epidemiology}.
\newblock {\em Statistical Science}, 31(1):96--118.

\bibitem[Fearnhead and Prangle, 2012]{Fearnhead2012}
Fearnhead, P. and Prangle, D. (2012).
\newblock {Constructing summary statistics for approximate Bayesian
  computation: Semi-automatic approximate Bayesian computation}.
\newblock {\em Journal of the Royal Statistical Society. Series B: Statistical
  Methodology}, 74(3):419--474.

\bibitem[Frazier et~al., 2019]{Frazier2019}
Frazier, D.~T., Maneesoonthorn, W., Martin, G.~M., and McCabe, B. P.~M. (2019).
\newblock Approximate {Bayesian} forecasting.
\newblock {\em International Journal of Forecasting}, 35(2):521--539.

\bibitem[Gelman et~al., 2013]{Gelman2013}
Gelman, A., Carlin, J.~B., Stern, H.~S., Dunson, D.~B., Vehtari, A., and Rubin,
  D.~B. (2013).
\newblock {\em {Bayesian data analysis}}.
\newblock Chapman \& Hall/CRC Texts in Statistical Science, third edition.

\bibitem[Gelman et~al., 1996]{Gelman1996}
Gelman, A., Meng, X.-L., and Stern, H. (1996).
\newblock Posterior predictive assessment of model fitness via realized
  discrepancies.
\newblock {\em Statistica Sinica}, 6(4):733--760.

\bibitem[Gillespie, 1977]{Gillespie1977}
Gillespie, D.~T. (1977).
\newblock Exact stochastic simulation of coupled chemical reactions.
\newblock {\em The Journal of Physical Chemistry}, 81(25):2340--2361.

\bibitem[Golightly and Sherlock, 2019]{Golightly2019}
Golightly, A. and Sherlock, C. (2019).
\newblock Efficient sampling of conditioned {Markov} jump processes.
\newblock {\em Statistics and Computing}, 29(5):1149–1163.

\bibitem[Golightly and Wilkinson, 2011]{Golightly2011}
Golightly, A. and Wilkinson, D.~J. (2011).
\newblock {Bayesian parameter inference for stochastic biochemical network
  models using particle Markov chain Monte Carlo}.
\newblock {\em Interface Focus}, 1(6):807–820.

\bibitem[Grazian and Fan, 2020]{Grazian2019}
Grazian, C. and Fan, Y. (2020).
\newblock {A review of Approximate Bayesian Computation methods via density
  estimation: inference for simulator-models}.
\newblock {\em WIREs Computational Statistics}, 12(4):e1486.

\bibitem[Hainy et~al., 2016]{Hainy2016}
Hainy, M., Drovandi, C.~C., and McGree, J.~M. (2016).
\newblock Likelihood-free extensions for {Bayesian} sequentially designed
  experiments.
\newblock In Kunert, J., M{\"u}ller, C.~H., and Atkinson, A.~C., editors, {\em
  mODa 11 - Advances in Model-Oriented Design and Analysis}, pages 153--161.

\bibitem[Heggland and Frigessi, 2004]{Heggland2004}
Heggland, K. and Frigessi, A. (2004).
\newblock Estimating functions in indirect inference.
\newblock {\em Journal of the Royal Statistical Society. Series B (Statistical
  Methodology)}, 66(2):447--462.

\bibitem[Jasra, 2015]{Jasra2015}
Jasra, A. (2015).
\newblock Approximate {Bayesian} computation for a class of time series models.
\newblock {\em International Statistical Review}, 83(3):405--435.

\bibitem[Jasra et~al., 2012]{Jasra2012}
Jasra, A., Singh, S., Martin, J., and McCoy, E. (2012).
\newblock {Filtering via approximate Bayesian computation}.
\newblock {\em Statistics and Computing}, 22:1223–1237.

\bibitem[Jiang et~al., 2018]{Jiang2018}
Jiang, B., Wu, T.-W., and Wong, W. (2018).
\newblock Approximate {B}ayesian computation with {Kullback-Leibler} divergence
  as data discrepancy.
\newblock In {\em Proceedings of the Twenty-First International Conference on
  Artificial Intelligence and Statistics}, pages 1711--1721.

\bibitem[Järvenpää et~al., 2019]{Jarvenpaa2019plos}
Järvenpää, M., Sater, M. R.~A., Lagoudas, G.~K., Blainey, P.~C., Miller,
  L.~G., McKinnell, J.~A., Huang, S.~S., Grad, Y.~H., and Marttinen, P. (2019).
\newblock A {Bayesian} model of acquisition and clearance of bacterial
  colonization incorporating within-host variation.
\newblock {\em PLOS Computational Biology}, 15(4):1--25.

\bibitem[Järvenpää et~al., 2020]{Jarvenpaa2020_babc}
Järvenpää, M., Vehtari, A., and Marttinen, P. (2020).
\newblock {Batch simulations and uncertainty quantification in Gaussian process
  surrogate approximate Bayesian computation}.
\newblock In {\em Proceedings of the 36th Conference on Uncertainty in
  Artificial Intelligence (UAI)}, pages 779--788.

\bibitem[Kleinegesse et~al., 2021]{Kleinegesse2021}
Kleinegesse, S., Drovandi, C., and Gutmann, M.~U. (2021).
\newblock {Sequential Bayesian Experimental Design for Implicit Models via
  Mutual Information}.
\newblock {\em Bayesian Analysis}, 16(3):773--802.

\bibitem[Krüger et~al., 2021]{Kruger2020}
Krüger, F., Lerch, S., Thorarinsdottir, T., and Gneiting, T. (2021).
\newblock {Predictive Inference Based on Markov Chain Monte Carlo Output}.
\newblock {\em International Statistical Review}, 89(2):274--301.

\bibitem[Kypraios et~al., 2017]{Kypraios2017}
Kypraios, T., Neal, P., and Prangle, D. (2017).
\newblock {A tutorial introduction to Bayesian inference for stochastic
  epidemic models using Approximate Bayesian Computation}.
\newblock {\em Mathematical Biosciences}, 287:42--53.

\bibitem[Lauritzen, 1974]{Lauritzen1974}
Lauritzen, S.~L. (1974).
\newblock Sufficiency, prediction and extreme models.
\newblock {\em Scandinavian Journal of Statistics}, 1(3):128--134.

\bibitem[Lewis et~al., 2021]{Lewis2021}
Lewis, J.~R., MacEachern, S.~N., and Lee, Y. (2021).
\newblock {Bayesian Restricted Likelihood Methods: Conditioning on Insufficient
  Statistics in Bayesian Regression (with Discussion)}.
\newblock {\em Bayesian Analysis}, 16(4):1393--2854.

\bibitem[Lintusaari et~al., 2017]{Lintusaari2016}
Lintusaari, J., Gutmann, M.~U., Dutta, R., Kaski, S., and Corander, J. (2017).
\newblock {Fundamentals and Recent Developments in Approximate Bayesian
  Computation.}
\newblock {\em Systematic biology}, 66(1):e66--e82.

\bibitem[Marin et~al., 2012]{Marin2012}
Marin, J.~M., Pudlo, P., Robert, C.~P., and Ryder, R.~J. (2012).
\newblock {Approximate Bayesian computational methods}.
\newblock {\em Statistics and Computing}, 22(6):1167--1180.

\bibitem[Marjoram et~al., 2003]{Marjoram2003}
Marjoram, P., Molitor, J., Plagnol, V., and Tavare, S. (2003).
\newblock {Markov chain Monte Carlo without likelihoods.}
\newblock {\em Proceedings of the National Academy of Sciences of the United
  States of America}, 100(26):15324--8.

\bibitem[Martin et~al., 2019]{Martin2019}
Martin, G.~M., McCabe, B. P.~M., Frazier, D.~T., Maneesoonthorn, W., and
  Robert, C.~P. (2019).
\newblock {Auxiliary Likelihood-Based Approximate Bayesian Computation in State
  Space Models}.
\newblock {\em Journal of Computational and Graphical Statistics},
  28(3):508--522.

\bibitem[Martin et~al., 2014]{Martin2014}
Martin, J.~S., Jasra, A., Singh, S.~S., Whiteley, N., Del~Moral, P., and McCoy,
  E. (2014).
\newblock {Approximate Bayesian Computation for Smoothing}.
\newblock {\em Stochastic Analysis and Applications}, 32(3):397--420.

\bibitem[McKinley et~al., 2009]{McKinley2009}
McKinley, T., Cook, A.~R., and Deardon, R. (2009).
\newblock Inference in epidemic models without likelihoods.
\newblock {\em The International Journal of Biostatistics}, 5(1).

\bibitem[Numminen et~al., 2013]{Numminen2013}
Numminen, E., Cheng, L., Gyllenberg, M., and Corander, J. (2013).
\newblock Estimating the transmission dynamics of streptococcus pneumoniae from
  strain prevalence data.
\newblock {\em Biometrics}, 69(3):748--757.

\bibitem[O'Hagan and Forster, 2004]{Ohagan2004}
O'Hagan, A. and Forster, J. (2004).
\newblock {\em {Advanced Theory of Statistics, Bayesian inference}}.
\newblock Arnold, London, UK, second edition.

\bibitem[Papamakarios and Murray, 2016]{Papamakarios2016}
Papamakarios, G. and Murray, I. (2016).
\newblock {Fast e-free inference of simulation models with Bayesian conditional
  density estimation}.
\newblock In {\em Advances in Neural Information Processing Systems 29}.

\bibitem[Papamakarios et~al., 2019]{Papamakarios2018}
Papamakarios, G., Sterratt, D., and Murray, I. (2019).
\newblock Sequential neural likelihood: Fast likelihood-free inference with
  autoregressive flows.
\newblock In {\em Proceedings of the 22nd International Conference on
  Artificial Intelligence and Statistics}, pages 837--848.

\bibitem[Pesonen et~al., 2021]{Pesonen2021}
Pesonen, H., Simola, U., Köhn-Luque, A., Vuollekoski, H., Lai, X., Frigessi,
  A., Kaski, S., Frazier, D.~T., Maneesoonthorn, W., Martin, G.~M., and
  Corander, J. (2021).
\newblock {ABC of the Future}.
\newblock Available at \url{https://arxiv.org/abs/2112.12841}.

\bibitem[Picchini, 2014]{Picchini2014}
Picchini, U. (2014).
\newblock {Inference for SDE Models via Approximate Bayesian Computation}.
\newblock {\em Journal of Computational and Graphical Statistics},
  23(4):1080--1100.

\bibitem[Prangle, 2017]{Prangle2017}
Prangle, D. (2017).
\newblock Adapting the {ABC} distance function.
\newblock {\em Bayesian Analysis}, 12(1):289--309.

\bibitem[Price et~al., 2018]{Price2018}
Price, L.~F., Drovandi, C.~C., Lee, A., and Nott, D.~J. (2018).
\newblock Bayesian synthetic likelihood.
\newblock {\em Journal of Computational and Graphical Statistics}, 27(1):1--11.

\bibitem[Pritchard et~al., 1999]{Pritchard1999}
Pritchard, J.~K., Seielstad, M.~T., Perez-Lezaun, A., and Feldman, M.~W.
  (1999).
\newblock {Population growth of human {Y} chromosomes: a study of {Y}
  chromosome microsatellites.}
\newblock {\em Molecular Biology and Evolution}, 16(12):1791--1798.

\bibitem[Ratmann et~al., 2009]{Ratmann2009}
Ratmann, O., Andrieu, C., Wiuf, C., and Richardson, S. (2009).
\newblock Model criticism based on likelihood-free inference, with an
  application to protein network evolution.
\newblock {\em Proceedings of the National Academy of Sciences of the United
  States of America}, 106(26):10576--10581.

\bibitem[Rudin, 1987]{Rudin1987}
Rudin, W. (1987).
\newblock {\em Real and complex analysis}.
\newblock McGraw-Hill, third edition.

\bibitem[Schervish, 1995]{Schervish1995}
Schervish, M.~J. (1995).
\newblock {\em Theory of statistics}.
\newblock Springer-Verlag.

\bibitem[Shestopaloff and Neil, 2014]{Shestopaloff2014}
Shestopaloff, A.~Y. and Neil, R.~M. (2014).
\newblock {On Bayesian inference for the M/G/1 queue with efficient MCMC
  sampling}.
\newblock Available at \url{https://arxiv.org/abs/1401.5548}.

\bibitem[Simola et~al., 2021]{Simola2021}
Simola, U., Cisewski-Kehe, J., Gutmann, M.~U., and Corander, J. (2021).
\newblock {Adaptive Approximate Bayesian Computation Tolerance Selection}.
\newblock {\em Bayesian Analysis}, 16(2):397--423.

\bibitem[Sisson et~al., 2019]{Sisson2019}
Sisson, S., Fan, Y., and Beaumont, M. (2019).
\newblock {\em {Handbook of Approximate Bayesian Computation}}.
\newblock {New York: Chapman and Hall/CRC}.

\bibitem[Sisson et~al., 2007]{Sisson2007}
Sisson, S.~A., Fan, Y., and Tanaka, M.~M. (2007).
\newblock {Sequential Monte Carlo without likelihoods.}
\newblock {\em Proceedings of the National Academy of Sciences of the United
  States of America}, 104(6):1760--5.

\bibitem[Skibinsky, 1967]{Skibinsky1967}
Skibinsky, M. (1967).
\newblock {Adequate Subfields and Sufficiency}.
\newblock {\em The Annals of Mathematical Statistics}, 38(1):155--161.

\bibitem[Stein and Shakarchi, 2005]{Stein2005}
Stein, E.~M. and Shakarchi, R. (2005).
\newblock {\em Real Analysis: Measure Theory, Integration, and {Hilbert}
  Spaces}.
\newblock Princeton University Press.

\bibitem[Tancredi, 2019]{Tancredi2019}
Tancredi, A. (2019).
\newblock {Approximate Bayesian inference for discretely observed
  continuous-time multi-state models}.
\newblock {\em Biometrics}, 75(3):966--977.

\bibitem[Tavar{\'e} et~al., 1997]{Tavare1997}
Tavar{\'e}, S., Balding, D.~J., Griffiths, R.~C., and Donnelly, P. (1997).
\newblock Inferring coalescence times from {DNA} sequence data.
\newblock {\em Genetics}, 145(2):505--518.

\bibitem[Thomas et~al., 2022]{Thomas2022}
Thomas, O., Dutta, R., Corander, J., Kaski, S., and Gutmann, M.~U. (2022).
\newblock {Likelihood-Free Inference by Ratio Estimation}.
\newblock {\em Bayesian Analysis}, 17(1):1--31.

\bibitem[Toni et~al., 2009]{Toni2009}
Toni, T., Welch, D., Strelkowa, N., Ipsen, A., and Stumpf, M. P.~H. (2009).
\newblock {Approximate Bayesian computation scheme for parameter inference and
  model selection in dynamical systems.}
\newblock {\em Journal of the Royal Society, Interface}, 6(31):187--202.

\bibitem[Vankov et~al., 2019]{Vankov2019}
Vankov, E.~R., Guindani, M., and Ensor, K.~B. (2019).
\newblock {Filtering and Estimation for a Class of Stochastic Volatility Models
  with Intractable Likelihoods}.
\newblock {\em Bayesian Analysis}, 14(1):29--52.

\bibitem[Vehtari and Ojanen, 2012]{Vehtari2012}
Vehtari, A. and Ojanen, J. (2012).
\newblock {A survey of Bayesian predictive methods for model assessment,
  selection and comparison}.
\newblock {\em Statistics Surveys}, 6:142--228.

\bibitem[Warne et~al., 2019]{Warne2019}
Warne, D.~J., Baker, R.~E., and Simpson, M.~J. (2019).
\newblock Simulation and inference algorithms for stochastic biochemical
  reaction networks: from basic concepts to state-of-the-art.
\newblock {\em Journal of The Royal Society Interface}, 16(151):20180943.

\bibitem[Warne et~al., 2020]{Warne2020}
Warne, D.~J., Baker, R.~E., and Simpson, M.~J. (2020).
\newblock A practical guide to pseudo-marginal methods for computational
  inference in systems biology.
\newblock {\em Journal of Theoretical Biology}, 496:110255.

\bibitem[Wilkinson, 2019]{Wilkinson2018}
Wilkinson, D.~J. (2019).
\newblock {\em Stochastic modelling for systems biology}.
\newblock Chapman \& Hall/CRC, third edition.

\bibitem[Wilkinson, 2013]{Wilkinson2013}
Wilkinson, R.~D. (2013).
\newblock {Approximate Bayesian computation (ABC) gives exact results under the
  assumption of model error}.
\newblock {\em Statistical Applications in Genetics and Molecular Biology},
  12(2):129--141.

\bibitem[Wood, 2010]{Wood2010}
Wood, S.~N. (2010).
\newblock {Statistical inference for noisy nonlinear ecological dynamic
  systems.}
\newblock {\em Nature}, 466:1102--1104.

\end{thebibliography}
\bibliographystyle{apalike}

\clearpage
\appendix

\setcounter{page}{1}
\setcounter{equation}{0}
\setcounter{footnote}{0}
\setcounter{figure}{0}
\renewcommand{\theequation}{\thesection.\arabic{equation}}
\renewcommand{\thefootnote}{\thesection.\arabic{footnote}}
\renewcommand{\thefigure}{\thesection.\arabic{figure}}

\section*{Supplementary materials for ``On predictive inference for intractable models via approximate Bayesian computation''}
\begin{center}
    \large
    Marko J\"{a}rvenp\"{a}\"{a} and Jukka Corander \\
    \today
\end{center}

{\color{\revcol}
\section{An alternative to \ABCP{}} \label{appe:abcpalt}

\ABCP{} is based on approximating $\pdf(\tl{y},y\cond\theta)$. One could also approximate $\pdf(\tl{y}\cond\theta,y)$ with
\begin{align}
    \pdfabc_{\tl{h}}(\tl{y}\cond\theta,\tl{\sss}_y) &
    \eqdef \frac{\int \tl{\K}_{\tl{h}}(\tl{\discr}(\tl{\sss}_y,\tl{\sss}_{z'})) \pdf(\tl{y},\tl{\sss}_{z'}\cond \theta) \ud \tl{\sss}_{z'}}{\iint \tl{\K}_{\tl{h}}(\tl{\discr}(\tl{\sss}_y,\tl{\sss}_{z'})) \pdf(\tl{y},\tl{\sss}_{z'}\cond \theta) \ud \tl{\sss}_{z'} \ud \tl{y}}
    \label{eq:abcpostpredfalt2}
\end{align}
for each $\theta$. (\ref{eq:abcpostpredfalt2}) is formed as the ``standard'' ABC posterior $\pdfabc_h(\theta\cond \sss_y)$ except that $\tl{y}$ appears in the place of $\theta$ and the parameter $\theta$ is conditioned on. This approximation can be useful when an accurate point estimate for $\theta$ is already available. This approximation is also used in Section \ref{subsec:lv}, given the specific parameter value of $\theta=\thetatrue$. The resulting approximate density is in fact exact (up to the sampling error) in the second data realization case of Section \ref{subsec:lvmis}.

An alternative to \ABCP{}, which also requires only the ability to jointly sample from $\pdf(\tl{y},y\cond\theta)$, is obtained by using (\ref{eq:abcpostpredfalt2}) in the \ABCF{} definition (\ref{eq:abcpostpredf}) so that 
\begin{align}
    \pdf(\tl{y}\cond y) 
    \approx
    \pdfabcfa_{\tl{h},h}(\tl{y}\cond \tl{\sss}_y; \sss_y) 
    &\eqdef \int \pdfabc_{\tl{h}}(\tl{y}\cond\theta,\tl{\sss}_y) \pdfabc_h(\theta\cond \sss_y) \ud\theta.
    \label{eq:abcpostpredfalt1}
\end{align}
We can also define the joint density\footnote{{\color{\revcol}We of course cannot directly manipulate this joint density using the basic fact $\pdf(a,b)=\pdf(a\cond b)\pdf(b)=\pdf(b\cond a)\pdf(a)$ (where $a$ and $b$ are random vectors) since $\pdfabc_{\tl{h}}(\tl{y}\cond\theta,\tl{\sss}_y)$ and $\pdfabc_h(\theta\cond \sss_y)$ are conditioned on different summary statistics. Similar observation also holds for \ABCF{}.}}
\begin{align}
\pdf(\tl{y},\theta \cond y) 
    \approx \pdfabcfa_{\tl{h},h}(\tl{y},\theta \cond \tl{\sss}_y; \sss_y) 
    \eqdef \pdfabc_{\tl{h}}(\tl{y}\cond\theta,\tl{\sss}_y) \pdfabc_h(\theta\cond \sss_y). 
    \label{eq:abcpostpredfalt3}
\end{align}
This approach is referred as \ABCFA{}. 
Note that a different threshold $\smash{\tl{h}}$, kernel $\smash{\tl{\K}}$, summary statistic $\tl{s}$ and discrepancy $\tl{\discr}$ can be used for the ``nested'' ABC approximation $\pdfabc_{\tl{h}}(\tl{y}\cond\theta,\tl{\sss}_y)$ as for $\pdfabc_h(\theta\cond \sss_y)$. 
The approximation $\pdfabc_{\tl{h}}(\tl{y}\cond\theta,\tl{\sss}_y)$ uses a separate set of pseudo-data which we have denoted by $z'$ to distinguish it from $z$ used in $\pdfabc_h(\theta\cond \sss_y)$. 

We can write the \ABCFA{} approximation as
\begin{align}
    \pdfabcfa_{\tl{h},h}(\tl{y},\theta \cond \tl{\sss}_y; \sss_y) 
    &= \pdfabc_{\tl{h}}(\tl{y}\cond\theta,\tl{\sss}_y) \pdfabc_h(\theta\cond \sss_y) \\
    &=
    \frac{\int \tl{\K}_{\tl{h}}(\tl{\discr}(\tl{\sss}_y,\tl{\sss}_{z'})) \pdf(\tl{y},\tl{\sss}_{z'}\cond \theta) \ud\tl{\sss}_{z'}}{\iint \tl{\K}_{\tl{h}}(\tl{\discr}(\tl{\sss}_y,\tl{\sss}_{z'})) \pdf(\tl{y},\tl{\sss}_{z'}\cond \theta) \ud \tl{\sss}_{z'} \ud \tl{y}}
    \frac{\int \K_{h}(\discr(\sss_y,\sss_{z})) \pdf(\sss_{z}\cond \theta) \pdf(\theta) \ud\sss_{z}}{\iint \K_{h}(\discr(\sss_y,\sss_{z})) \pdf(\sss_{z}\cond \theta) \pdf(\theta) \ud\sss_{z} \ud\theta} \\
    &= \frac{\int \tl{\K}_{\tl{h}}(\tl{\discr}(\tl{\sss}_y,\tl{\sss}_{z'})) \pdf(\tl{y},\tl{\sss}_{z'}\cond \theta) \pdf(\theta) \ud\tl{\sss}_{z'}}{\iiint \K_{h}(\discr(\sss_y,\sss_{z})) \pdf(\tl{y},\sss_{z}\cond \theta) \pdf(\theta) \ud\sss_{z} \ud\theta \ud\tl{y}}
    \frac{\pdfabc_{h}(\sss_y\cond\theta)}{\pdfabc_{\tl{h}}(\tl{\sss}_y\cond\theta)},
    \label{eq:abcpostpredfalt4}
\end{align}
where $\pdfabc_{h}(\sss_y\cond\theta)$ and $\pdfabc_{\tl{h}}(\tl{\sss}_y\cond\theta)$ denote ABC likelihoods, see Section \ref{subsec:abc}. 
Now, if $\tl{\sss}=\sss$, $\tl{h}=h$, $\tl{K}=K$ and $\tl{\discr}=\discr$, then $\pdfabc_{h}(\sss_y\cond\theta) = \pdfabc_{\tl{h}}(\tl{\sss}_y\cond\theta)$ so that these terms cancel out in (\ref{eq:abcpostpredfalt4}) and the resulting formula equals (\ref{eq:abcjointpostpred}). This shows that \ABCFA{} defines the same target approximation as \ABCP{} in this particular case. Hence, \ABCFA{} could be considered as a generalization of \ABCP{}. 

\ABCFA{} does not seem to provide evident, general advantages over its special case \ABCP{}.
For example, matching of predictive sufficient summary statistics is required in both cases. Specifically, $\tl{\sss}$ needs to ideally satisfy the condition (\ref{eq:predsuffstrong2}) while $\sss$ needs to be parametric sufficient satisfying (\ref{eq:predsuffstrong1}). Also, it is not immediately clear how to most efficiently sample from (\ref{eq:abcpostpredfalt1}) as the normalization constant of $\pdfabc_{\tl{h}}(\tl{y}\cond\theta,\tl{\sss}_y)$ depends on $\theta$ or how to determine $\tl{h}$ which in principle could depend on $\theta$. On the other hand, \ABCFA{} might facilitate more convenient implementation than \ABCP{} in a sense that the estimation of $\theta$ and $\tl{y}$ can be separated. More detailed investigation is however left as a potential topic for future work. 
}

\section{Additional mathematical details} \label{appe:technicaldetails}

\subsection{Proof of Proposition \ref{thm:abcp}} \label{appe:theoryproof}

We obtain
\begin{align}
    \lim_{t\rightarrow\infty} \pdfabcp_{h_t}(\tl{y}\cond\sss_y) 
    &= \lim_{t\rightarrow\infty} \frac{\iint\indic_{\sss_z\in\A_t}\pdf(\tl{y},\sss_z\cond\theta)\pdf(\theta)\ud\sss_z\ud\theta}{\iiint\indic_{\sss_z\in\A_t}\pdf(\tl{y},\sss_z\cond\theta)\pdf(\theta)\ud\sss_z\ud\theta\ud\tl{y}} \\
    &= \lim_{t\rightarrow\infty} \frac{\iint\indic_{\sss_z\in\A_t}\pdf(\tl{y},\sss_z,\theta)\ud\theta\ud\sss_z}{\iiint\indic_{\sss_z\in\A_t}\pdf(\tl{y},\sss_z,\theta)\ud\theta\ud\tl{y}\ud\sss_z} \\
    &= \lim_{t\rightarrow\infty} \frac{\int_{\sss_z\in\A_t}\pdf(\tl{y},\sss_z)\ud\sss_z}{\int_{\sss_z\in\A_t}\pdf(\sss_z)\ud\sss_z} \\
    &= \frac{\lim_{t\rightarrow\infty}\frac{1}{|\A_t|}\int_{\sss_z\in\A_t}\pdf(\tl{y},\sss_z)\ud\sss_z}{\lim_{t\rightarrow\infty}\frac{1}{|\A_t|}\int_{\sss_z\in\A_t}\pdf(\sss_z)\ud\sss_z} \\
    &= \frac{\pdf(\tl{y},\sss_y)}{\pdf(\sss_y)} \\
    &= \pdf(\tl{y}\cond\sss_y),
\end{align}
where on the second line we have used Tonelli's theorem to change the order of integration and where the fifth equality holds almost everywhere and follows from Lebesgue differentiation theorem (see e.g.~\citet[Chapter~7]{Rudin1987} or \citet[Chapter 3]{Stein2005}) which requires \ref*{it:a4} and \ref*{it:a5}. 

Similar reasoning as above further shows that 
\begin{align}
    \lim_{t\rightarrow\infty} \pdfabcp_{h_t}(\tl{y},\theta\cond\sss_y) 
    &= \lim_{t\rightarrow\infty} \frac{\int\indic_{\sss_z\in\A_t}\pdf(\tl{y},\sss_z\cond\theta)\pdf(\theta)\ud\sss_z}{\iiint\indic_{\sss_z\in\A_t}\pdf(\tl{y},\sss_z\cond\theta)\pdf(\theta)\ud\sss_z\ud\theta\ud\tl{y}} \\
    &= \lim_{t\rightarrow\infty} \frac{\int_{\sss_z\in\A_t}\pdf(\tl{y},\theta,\sss_z)\ud\sss_z}{\int_{\sss_z\in\A_t}\pdf(\sss_z)\ud\sss_z} \\
    &= \frac{\lim_{t\rightarrow\infty}\frac{1}{|\A_t|}\int_{\sss_z\in\A_t}\pdf(\tl{y},\theta,\sss_z)\ud\sss_z}{\lim_{t\rightarrow\infty}\frac{1}{|\A_t|}\int_{\sss_z\in\A_t}\pdf(\sss_z)\ud\sss_z} \\
    &= \frac{\pdf(\tl{y},\theta,\sss_y)}{\pdf(\sss_y)} \\
    &= \pdf(\tl{y},\theta\cond\sss_y),
\end{align}
where the fourth equality holds almost everywhere as in the previous case.

\subsection{Mathematical details related to Example \ref{ex:markovmodel}} \label{appe:markovdetails}

We justify the various results in Example \ref{ex:markovmodel}. Filling the details of some straightforward computations are left for the reader. First, we compute
\begin{align}
    \pdf(c\cond y,\teta,\sigma^2) &\propto {\prod_{i=1}^n\pdf(y_i\cond y_{i-1},c,\teta,\sigma^2)\pdf(c)} 
    \propto \prod_{i=1}^n \Normal(y_i\cond c + \teta y_{i-1},\sigma^2) \\
    &\propto \e^{-\frac{1}{2\sigma^2}(y_1-c)^2} \prod_{i=2}^n \e^{-\frac{1}{2\sigma^2}(x_i-c-\phi y_{i-1})^2} 
    \propto \e^{-\frac{1}{2\sigma^2}[nc^2 - 2c(\sum_{i=1}^n y_i) - \teta\sum_{i=1}^{n-1}y_i]} \\
    &\propto \Normal(c\cond \baryteta, \sigma^2/n), \label{eq:markovpostderiv}
\end{align}
which shows $\baryteta$ is parametric sufficient. Since $\pdf(y_{n+p}\cond y_{1:n},\theta) = \pdf(y_{n+p}\cond y_n,\theta)$ for any $p\geq 1$ by the Markov property, $(\baryteta,y_n)$ is predictive sufficient for any $y_{n+p}$, and in particular for $y_{n+1}$. 

In the rest of this section\footnote{The following results in fact hold more generally, e.g.~(\ref{eq:markovmean}) when $\teta\neq 1$ and (\ref{eq:markovvar}) when $|\teta|\neq 1$.} we assume $|\teta| < 1$.  
First we notice that the Markov process can be written as
\begin{equation}
    y_t = c\frac{1-\teta^t}{1-\teta} + \sum_{i=1}^t \teta^{t-i}\epsilon_i, \quad t=0,1,\ldots
    \label{eq:myt}
\end{equation}
where we have used the geometric sum formula which holds because $\teta\neq 1$. Using (\ref{eq:myt}), the basic properties of expectation and (co)variance, and the fact that $\epsilon_i$ are \iid{}, one can show that\footnote{When $|\teta|<1$ and if we let $t\rightarrow\infty$, we obtain the well-known formulas for the mean and variance of the stationary order one autoregressive model AR(1).}
\begin{align}
    \mean(y_t\cond\theta) &= c\frac{1-\teta^t}{1-\teta}, \label{eq:markovmean} \\
    \Var(y_t\cond\theta) &= \sigma^2\frac{1-\teta^{2t}}{1-\teta^2}, \label{eq:markovvar} \\
    \cov(y_s,y_t\cond\theta) &= \sigma^2\teta^{|s-t|}\frac{1-\teta^{2\min\{s,t\}}}{1-\teta^2} \label{eq:markovcov}
\end{align}
for any $s,t=1,2,\ldots$. 
Recall that $\baryteta \eqdef ({(1-\teta)\sum_{i=1}^{n-1}y_i + y_n})/{n}$. Since $y_{1:n}$ are clearly jointly Gaussian and since $\baryteta$ depends linearly on $y_{1:n}$, $\baryteta$ also follows Gaussian distribution whose mean and variance can be easily shown to be
\begin{align}
    \mean(\baryteta\cond\theta) &= c, \label{eq:markovsummarymean} \\
    \Var(\baryteta\cond\theta) &= \sigma^2/n. \label{eq:markovsummaryvar}
\end{align}
Next, using (\ref{eq:markovsummarymean}), (\ref{eq:markovsummaryvar}) and a standard Gaussian identity, we obtain (\ref{eq:ex1:cpost}):
\begin{equation}
    \pdfabc_h(c \cond \baryteta) 
    \propto \int_{-\infty}^{\infty} \Normal(\baryteta \cond \bar{z}_\teta, h^2) \Normal(\bar{z}_\teta \cond c, \sigma^2/n) \pi(c) \ud \bar{z}_\teta
    \propto \Normal(c \cond \baryteta, \sigma^2/n + h^2). \label{eq:csumpost}
\end{equation}

Using (\ref{eq:markovmean}), (\ref{eq:markovvar}) and (\ref{eq:markovcov}) we obtain 
\begin{equation}
    \begin{bmatrix} y_{n+p} \\ y_n \end{bmatrix} \,\bigg|\, \theta 
    \sim \Normal_2\left( \frac{c}{1-\teta}\begin{bmatrix} {1-\teta^{n+p}}{} \\ {1-\teta^{n}}{} \end{bmatrix}, 
    \frac{\sigma^2}{1-\teta^2} \begin{bmatrix} {1-\teta^{2(n+p)}}{} & {\teta^p(1-\teta^{2n})}{} \\ {\teta^p(1-\teta^{2n})}{} & {1-\teta^{2n}}{} \end{bmatrix} \right).
\end{equation}
Using a well-known formula for conditional Gaussian density further produces
\begin{equation}
    y_{n+p} \cond y_n, \theta \sim \Normal\left(c\frac{1-\teta^p}{1-\teta} + \teta^p y_n, \sigma^2\frac{1-\teta^{2p}}{1-\teta^2}\right). 
    \label{eq:gausynp}
\end{equation}
Using $\pdf(y_{n+p}\cond y_{1:n},\theta) = \pdf(y_{n+p}\cond y_n,\theta)$, (\ref{eq:gausynp}) and (\ref{eq:csumpost}) we obtain 
\begin{align}
    \pdfabcf_{h}(y_{n+p} \cond y; \baryteta)
    &= \int_{-\infty}^{\infty} \pdf(y_{n+p}\cond y,\theta) \pdfabc_h(c \cond \baryteta) \ud c \\
    &= \Normal\left(y_{n+p} \mcond \ringy_{\teta,p}, \sigma^2\frac{1-\teta^{2p}}{1-\teta^2} + \frac{(1-\teta^p)^2}{(1-\teta)^2}(\sigma^2/n + h^2) \right), \quad \ringy_{\teta,p} \eqdef \frac{1-\teta^p}{1-\teta}\baryteta + \teta^p y_n, 
    \label{eq:postpredextrap}
\end{align}
where the integral is computed by deducing that the result is Gaussian density whose mean (variance) follows by using the law of total expectation (variance) or, alternatively, by applying Gaussian identities. 
The result (\ref{eq:ex1:postpredf}) follows immediately by setting $p=1$. 
Similarly, the discussion below (\ref{eq:markovpostderiv}) and the result (\ref{eq:postpredextrap}) with $p=1$ and $h=0$ imply (\ref{eq:ex1:postpredtrue}). As the resulting formula depends on data $y_{1:n}$ only via $\ringyteta$, the summary statistic $\ringyteta$ is predictive sufficient (in the weaker sense). 

We then justify (\ref{eq:ex1:postprednonopt}) and (\ref{eq:ex1:postpredopt}). As the latter result follows by similar computations but with $\ringyteta$ in place of $\baryteta$, we only outline the derivation of the former result. Using (\ref{eq:markovcov}) and some straightforward computations we obtain
\begin{equation}
    \cov(y_{n+1},\baryteta \cond \theta) = \sigma^2\frac{\teta(1-\teta^{n})}{n(1-\teta)}. \label{eq:markovsumycov}
\end{equation}
The joint distribution of $y_{1:n+1}=(y_1,\ldots,y_{n+1})$ is clearly Gaussian and so is that of $(y_{1:n+1},\baryteta)$. By using (\ref{eq:markovmean}), (\ref{eq:markovvar}), (\ref{eq:markovsummarymean}), (\ref{eq:markovsummaryvar}) and (\ref{eq:markovsumycov}), we further obtain
\begin{equation}
    \begin{bmatrix} y_{n+1} \\ \baryteta \end{bmatrix} \,\bigg|\, \theta 
    \sim \Normal_2\left( c\begin{bmatrix} \frac{1-\teta^{n+1}}{1-\teta} \\ 1 \end{bmatrix}, 
    \sigma^2 \begin{bmatrix} \frac{1-\teta^{2(n+1)}}{1-\teta^2} & \frac{\teta(1-\teta^{n})}{n(1-\teta)} \\ \frac{\teta(1-\teta^{n})}{n(1-\teta)} & \frac{1}{n} \end{bmatrix}  \right).
\end{equation}
This implies 
\begin{equation}
    \pdf( y_{n+1} \cond \baryteta, \theta) 
    = \Normal\left(y_{n+1} \mcond c + \frac{\baryteta\teta(1-\teta^n)}{1-\teta}, \sigma^2\left( \frac{1-\teta^{2(n+1)}}{1-\teta^2} - \frac{\teta^2(1-\teta^n)^2}{n(1-\teta)^2} \right)\!\right). \label{eq:markovpred2}
\end{equation}
Equation (\ref{eq:ex1:postprednonopt}) then follows as 
\begin{align}
    \pdfabcp_{h}(y_{n+1} \cond \baryteta) 
    &\propto \iint \Normal(\baryteta\cond\barz_\teta,h^2) \pdf(y_{n+1}\cond\barz_\teta,c) \Normal(\barz_\teta\cond c, \sigma^2/n) \pdf(c) \ud\barz_\teta \ud c \\
    &\propto \int \Normal(\baryteta\cond\barz_\teta,h^2) \int \Normal\left(y_{n+1}\mcond c + \frac{\baryteta\teta(1-\teta^n)}{1-\teta} ,a_{\teta, n}\sigma^2\right) \Normal(c \cond \barz_\teta, \sigma^2/n) \ud c \ud\barz_\teta \\
    &= \int \Normal\left(y_{n+1} \mcond b_{\teta, n}\barz_\teta, a_{\teta, n}\sigma^2 + \sigma^2/n \right) \Normal(\barz_\teta \cond \baryteta,h^2) \ud\barz_\teta \\
    &= \Normal\left(y_{n+1} \mcond b_{\teta, n}\baryteta, a_{\teta, n}\sigma^2 + \sigma^2/n + b^2_{\teta, n}h^2\right),
\end{align}
where on the second line we used (\ref{eq:markovpred2}).

It remains to verify the inequality in (\ref{eq:ex1:ab}):
\begin{align}
    a_{n,\phi} - 1 = \frac{1-\teta^{2(n+1)}}{1-\teta^2} - \frac{\teta^2(1-\teta^n)^2}{n(1-\teta)^2} - 1 
    &= \sum_{i=0}^n \teta^{2i} - \frac{\teta^2}{n}\bigg( \sum_{i=0}^{n-1} \teta^{i} \bigg)^{\!2} -1 \label{eq:polyn} \\
    &= \teta^2 \left( \sum_{i=0}^{n-1} \teta^{2i} - \frac{1}{n}\bigg( \sum_{i=0}^{n-1} \teta^{i} \bigg)^{\!2} \right) \\
    &\geq 0,
\end{align}
where we have used the Cauchy-Schwarz inequality. Since (\ref{eq:polyn}) is a polynomial of order $2n$ there are at most $n$ global minima where the equality holds (clearly $\teta=0$ and $\teta=1$ are such points) and elsewhere the inequality is strict.

\subsection{Justification for Equation (\ref{eq:predsuffacc})} \label{appe:predsuffacc}

Suppose first $|\teta| > 1$. By using (\ref{eq:markovmean}) and (\ref{eq:markovvar}) we obtain
\begin{align}
    \prob(|z_n-y_n|\leq \tl{h}\cond c) 
    &= \frac{1}{\sqrt{2\pi\sigma^2}}\sqrt{\frac{1-\teta^2}{1-\teta^{2n}}} \underbrace{\int_{y_n-\tl{h}}^{y_n+\tl{h}} \e^{-\frac{1}{2\sigma^2}\frac{1-\teta^2}{1-\teta^{2n}}\big( z_n - c\frac{1-\teta^n}{1-\teta} \big)^2} \ud z_n}_{\leq 2\tl{h}} 
    \label{eq:c1}
    \\
    &\leq \frac{\sqrt{2}\tl{h}}{\sqrt{\pi\sigma^2}}{\frac{\sqrt{\teta^2-1}}{\sqrt{\teta^{2n}-1}}}.
    \label{eq:c2}
\end{align}
The maximum of (\ref{eq:c1}) exists and is clearly obtained when $c\frac{1-\teta^n}{1-\teta}=y_n$ so that (\ref{eq:c2}) gives its upper bound.

Suppose $|\teta| < 1$. The maximum wrt.~$c$ is again obtained when $c\frac{1-\teta^n}{1-\teta}=y_n$ and we then compute
\begin{align}
    \max_{c\in\reals}\prob(|z_n-y_n|\leq \tl{h}\cond c) 
    &= \frac{1}{\sqrt{2\pi\sigma^2}}\sqrt{\frac{1-\teta^2}{1-\teta^{2n}}} \max_{c\in\reals} \int_{y_n-\tl{h}}^{y_n+\tl{h}} \e^{-\frac{1}{2\sigma^2}\frac{1-\teta^2}{1-\teta^{2n}}\big( z_n - c\frac{1-\teta^n}{1-\teta} \big)^2} \ud z_n \\
    &= \frac{1}{\sqrt{2\pi\sigma^2}}\sqrt{\frac{1-\teta^2}{1-\teta^{2n}}} \underbrace{\int_{y_n-\tl{h}}^{y_n+\tl{h}} \e^{-\frac{1}{2\sigma^2}\frac{1-\teta^2}{1-\teta^{2n}}\big( z_n - y_n \big)^2} \ud z_n}_{\geq 2\tl{h}\e^{-\frac{1}{2\sigma^2}\frac{1-\teta^2}{1-\teta^{2n}}\tl{h}^2}} \\
    &\geq \frac{\sqrt{2}\tl{h}}{\sqrt{\pi\sigma^2}}\sqrt{\frac{1-\teta^2}{1-\teta^{2n}}} \e^{-\frac{1}{2\sigma^2}\frac{1-\teta^2}{1-\teta^{2n}}\tl{h}^2} \\
    &\geq \frac{\sqrt{2}\tl{h}}{\sqrt{\pi\sigma^2}}\sqrt{1-\teta^2}\e^{-\tl{h}^2/(2\sigma^2)}.
\end{align}
%

Suppose now $\teta=-1$. We obtain $\mean(y_t\cond c) = c({1-(-1)^t})/2$ directly from (\ref{eq:markovmean}). By using (\ref{eq:myt}) we additionally compute $\Var(y_t\cond\theta) = t\sigma^2$ which holds for all $t=1,2,\ldots$. The result now follows similarly as when $|\teta|>1$ except that for odd $n$ the maximum wrt.~$c$ is obtained when $c=y_n$ and for even $n$ we interestingly have $\mean(y_n\cond c)=0$ so that in this case $\prob(|z_n-y_n|\leq \tl{h}\cond c)$ is constant wrt.~$c$. In both cases the maximum hence exists and the same upper bound is obtained. 
For the remaining case $\teta=1$ we compute $\mean(y_t\cond c)=tc$ and $\Var(y_t\cond\theta) = t\sigma^2$ which hold for all $t=1,2,\ldots$. The rest again follows similarly as when $|\teta|>1$.

\subsection{Mathematical details related to Example \ref{ex:statespacemodel}} \label{appe:ex:statespacemodeldetails}

We first notice that 
\begin{equation}
    \begin{bmatrix} y_{1:n} \\ v_{1:n} \end{bmatrix} \,\bigg|\, c 
    \sim \Normal_{2n}\left( c\begin{bmatrix} \mu \\ \mu \end{bmatrix}, 
    \begin{bmatrix} W & \Sigma \\ \Sigma & \Sigma \end{bmatrix}  \right),
    \label{eq:yvjoint}
\end{equation}
where we have $\mu_t = (1-\teta^t)/(1-\teta)$ by (\ref{eq:markovmean}) and where $W=\Sigma + \omega^2\Id$ with $\Sigma_{s,t} = \cov(v_s,v_t\cond c)$ by (\ref{eq:markovcov}). Note that $\mu,\Sigma$ and $W$ do not depend on parameter c.  
We use $\Sigma_{t:}$ to denote the row vector that contains the $t$th row of $\Sigma$ and similarly $\Sigma_{:t}$ is the $t$th column vector. It follows that
\begin{equation}
    v_n\cond y_{1:n},c \sim 
    \Normal(c(\mu_n - \Sigma_{n:}W^{-1}\mu) + \Sigma_{n:}W^{-1}y_{1:n}, \Sigma_{n,n}-\Sigma_{n:}W^{-1}\Sigma_{:n}), 
    \label{eq:ssfiltering}
\end{equation}
which is the ``batch'' solution for the filtering problem. Using $\pdf(y_{1:n}\cond c) = \Normal(y_{1:n}\cond c\mu, W)$, which follows from (\ref{eq:yvjoint}), and some straightforward computations (or, alternatively, by existing formulas for this linear Gaussian inference problem), we see that 
\begin{equation}
    c\cond y_{1:n}\sim\Normal((\mu\T W^{-1}\mu)^{-1}\underbrace{\mu\T W^{-1}y_{1:n}}_{s^{(1)}(y)}, (\mu\T W^{-1}\mu)^{-1}).
    \label{eq:sscpost}
\end{equation}
%

Now that we have derived (\ref{eq:ssfiltering}) and (\ref{eq:sscpost}), we can further obtain
\begin{align}
    \pdf(y_{n+1}\cond y_{1:n}) 
    &= \iiint \Normal(y_{n+1}\cond v_{n+1},\omega^2) \Normal(v_{n+1}\cond c+\teta v_n,\sigma^2) \pdf(v_n\cond y_{1:n},c) \pdf(c\cond y_{1:n}) \ud c \ud v_n \ud v_{n+1} \\
    &= \int \Normal(y_{n+1}\cond v_{n+1},\omega^2) \iint \Normal(v_{n+1}\cond c+\teta v_n,\sigma^2) \pdf(v_n\cond y_{1:n},c) \ud v_n \pdf(c\cond y_{1:n}) \ud c \ud v_{n+1} \label{eq:sspredpost1} \\
    \begin{split}
    &=\Normal\Big(
    \teta\underbrace{\Sigma_{n:} W^{-1}y_{1:n}}_{s^{(2)}(y)} + \,(1+\teta\mu_n-\teta\Sigma_{n:}W^{-1}\mu)(\mu\T W^{-1}\mu)^{-1}\underbrace{\mu\T W^{-1}y_{1:n}}_{s^{(1)}(y)}, \\
    &\myquad\quad \omega^2 + \sigma^2 + \teta^2(\Sigma_{n,n}-\Sigma_{n:}W^{-1}\Sigma_{:n}) + (1+\teta\mu_n-\teta\Sigma_{n:}W^{-1}\mu)^2(\mu\T W^{-1}\mu)^{-1}\Big),
    \end{split}
    \label{eq:ssmlinsuffstat}
\end{align}
where the integrals in (\ref{eq:sspredpost1}) are computed in a straightforward manner by using the laws of total expectation and variance. 

Finally, we use (\ref{eq:ssfiltering}) and (\ref{eq:sscpost}) to obtain
\begin{align}
    \begin{split}
    v_n\cond y_{1:n} \sim \Normal\Big( 
    &(\mu\T W^{-1}\mu)^{-1}\underbrace{\mu\T W^{-1}y_{1:n}}_{s^{(1)}(y)}(\mu_n - \Sigma_{n:}W^{-1}\mu) + \underbrace{\Sigma_{n:} W^{-1}y_{1:n}}_{s^{(2)}(y)}, \\
    &\Sigma_{n,n}-\Sigma_{n:}W^{-1}\Sigma_{:n} + (\mu_n - \Sigma_{n:}W^{-1}\mu)^2(\mu\T W^{-1}\mu)^{-1} \Big).
    \end{split}
    \label{eq:ssvnpost}
\end{align}
The joint Gaussian density of $(v_n,c)$ given $y_{1:n}$ can be formed using (\ref{eq:ssvnpost}), (\ref{eq:sscpost}) and the fact $\cov(v_n,c\cond y_{1:n}) = (\mu_n - \Sigma_{n:}W^{-1}\mu)(\mu\T W^{-1}\mu)^{-1}$.

\section{Additional results} \label{appe:experiments}

\subsection{Additional illustration in the case of Example \ref{ex:markovmodel}}

\begin{figure}[htbp]
\centering
\includegraphics[width=0.85\textwidth]{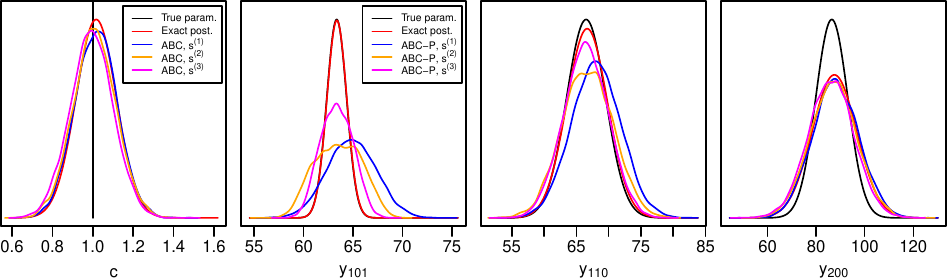}
\caption{Illustration of the effect of summary statistics $s^{(1)}(y)=\baryteta$, $s^{(2)}(y)=(\baryteta,y_n)$ and $s^{(3)}(y)=\ringyteta=\baryteta+\phi y_n$ on the ABC approximation accuracy. This example is the same as in Figure~\ref{fig:markov1} of the main paper except that here we used $\teta=0.99$. \emph{The first plot} on the left shows the ABC(-P/F) posterior for $c$ and \emph{the three other plots} the \ABCP{} posterior predictive distribution at some future time points.} \label{fig:markov2}
\end{figure}

We consider an illustration that is otherwise the same as the one in Example \ref{ex:markovmodel} except that we use $\teta=0.99$. 
The resulting posterior approximations are shown in Figure \ref{fig:markov2}. The results agree with the theoretical considerations. Interestingly, all summary statistics produce accurate posterior of $c$ in this particular case. The \ABCF{} posterior predictive densities are not shown for clarity and because they all were essentially the same as the exact posterior. This is not surprising because the ABC posterior for $c$ is accurate in all three cases. 

\begin{figure}[htbp] 
\centering
\begin{subfigure}{0.7\textwidth}
\centering
\includegraphics[width=\textwidth]{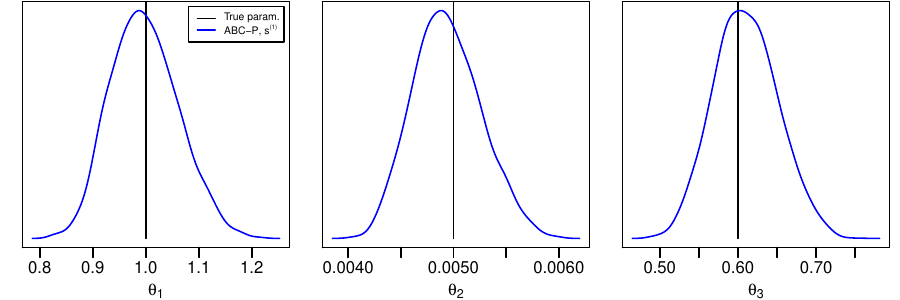}
\end{subfigure}
\\[0.2cm]
\begin{subfigure}{0.7\textwidth}
\centering
\includegraphics[width=\textwidth]{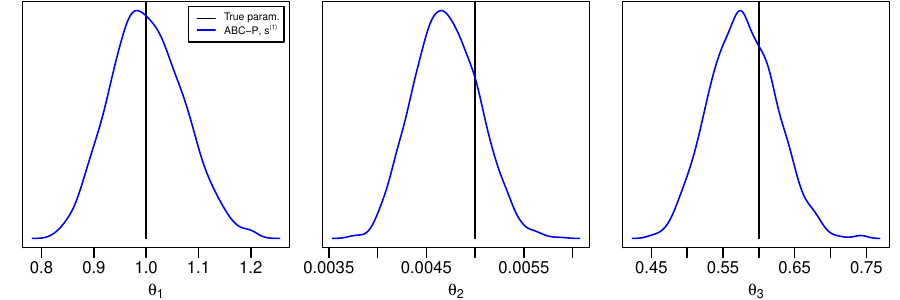}
\end{subfigure}
\caption{{\color{\revcol}Posterior distributions for the parameters of the Lotka-Volterra experiments corresponding to Figure~\ref{fig:lvmis1} in Section \ref{subsec:lvmis} of the main paper. \emph{Top row:} The first realization of data. \emph{Bottom row:} The second realization of data where the populations have become extinct. The black vertical line shows the true value of the parameter.}} \label{fig:lvmis2}
\end{figure}

{\color{\revcol}
\subsection{Computed ABC posteriors for the missing data case of Section \ref{subsec:lvmis}}

Finally, Figure \ref{fig:lvmis2} shows the computed ABC posterior distributions for the parameters of the Lotka-Volterra experiments of Section \ref{subsec:lvmis}. We can see that both data realizations lead to fairly similar ABC posteriors in this case.} 

\clearpage 

\end{document}